\documentclass[aps,prb,preprint]{revtex4-1}
\usepackage{graphicx}
\usepackage{epstopdf}

\usepackage{bm,amsmath,amssymb} 
\usepackage{color, soul} 
\usepackage{dcolumn}   
\usepackage{multirow}

\newcommand{\etal}{{\em et al}.\ }
\newcommand{\eg}{{\em e.g.,}}
\begin{document}

\title{First-Principles Study of Two-Dimensional Ferroelectrics Using Self-Consistent Hubbard Parameters}

\author{Jiawei Huang}
\affiliation{Zhejiang University, Hangzhou Zhejiang 310058, P.R. China}
\affiliation{School of Science, Westlake University, Hangzhou, Zhejiang 310024, China}
\author{Sang-Hoon Lee}
\affiliation{Korea Institute for Advanced Study, Seoul 02455, Korea}
\author{Andrew Supka}
\affiliation{Department of Physics and Science of Advanced Materials Program, Central Michigan University, Mt. Pleasant, Michigan 48859, United States}
\author{Young-Woo Son}
\affiliation{Korea Institute for Advanced Study, Seoul 02455, Korea}
\author{Shi Liu}
\email{liushi@westlake.edu.cn}
\affiliation{School of Science, Westlake University, Hangzhou, Zhejiang 310024, China}
\affiliation{Institute of Natural Sciences, Westlake Institute for Advanced Study,Hangzhou, Zhejiang 310024, China}

\date{\today}

\begin{abstract}{ The discovery of two-dimensional (2D) materials possessing switchable spontaneous polarization with atomic thickness opens up exciting opportunities to realize ultrathin, high-density electronic devices with potential applications ranging from memories and sensors to photocatalysis and solar cells.  First-principles methods based on density functional theory (DFT) have facilitated the discovery and design of 2D ferroelectrics (FEs). However, DFT calculations employing local and semilocal exchange-correlation functionals failed to predict accurately the band gaps for this family of low dimensional materials. Here, we present a DFT+$U$+$V$ study on 2D FEs represented by $\alpha$-In$_2$Se$_3$ and its homologous III$_2$-VI$_3$ compounds with both out-of-plane and in-plane polarization, using Hubbard parameters computed from first-principles. We find that ACBN0, a pseudo-hybrid density functional that allows self-consistent determination of $U$ parameters, improves the prediction of band gaps for all investigated 2D FEs with a computational cost much lower than the Heyd-Scuseria-Ernzerhof hybrid density functional. The inter-site Coulomb interaction $V$ becomes critical for accurate descriptions of the electronic structures of van der Waals heterostructures such as bilayer In$_2$Se$_3$ and In$_2$Se$_3$/InTe. Pertinent to the study of FE-based catalysis, we find that the application of self-consistent $U$ corrections can strongly affect the adsorption energies of open-shell molecules on the polar surfaces of 2D FEs. 
}
\end{abstract}

\maketitle

\newpage
\section{Introduction}
Ferroelectrics (FEs) with tunable electric polarization are technologically important functional materials used in a wide range of applications such as non-volatile memories, field effect transistors (FETs), sensors, and solar cells~\cite{Setter06p051606,Scott07p954,Huang18p1700560,Mikolajick20p1434}. To realize high-density electronic devices, it is essential for a ferroelectric to maintain robust room-temperature polarization at the nanoscale. However, conventional perovskite ferroelectrics such as Pb(Zr, Ti)O$_3$ suffer from the finite size effect: the out-of-plane polarization will disappear when the film thickness is below a critical value of a few nanometers due to the depolarization field arising from an incomplete screening of surface charges. This becomes a major obstacle for the scaling of ferroelectric-based electronic devices. For example, the first commercial ferroelectric random-access memory appeared in the early 1990s~\cite{Bondurant90p273}, but current state-of-art technology node remains to be 130~nm because a thick perovskite layer ($\approx$70~nm) is needed to maintain the polarization~\cite{McAdams04p667}. Developing ultrathin ferroelectrics with large switchable polarization at room temperature is an actively pursued goal. 

Two-dimensional (2D) materials with atomic thickness possessing spontaneous switchable polarization offers a potential solution to the scaling issue of ferroelectrics. The existence of 2D ferroelectricity was predicted more than fifty years ago by Onsager using a 2D Ising model~\cite{Onsager44p117}. More recently, first-principles methods especially density functional theory (DFT) calculations have played an important role in advancing the development of 2D ferroelectrics, successfully predicting a few 2D materials with ferroelectric polarization. For example, graphene-based materials functionalized with hydroxyl groups were predicted to be ferroelectric by Wu {\em et al.} based on DFT calculations~\cite{Wu13p081406l}. Shirodkar and Waghmare demonstrated with Landau theory analysis and first-principles calculations that the  $K_3$ mode of the centrosymmetric $1T$ ($c1T$) structure of MoS$_2$ monolayer could lead to the trimerization of Mo atoms, resulting in a distorted $1T$ ($d1T)$ phase with a spontanenous polarization of 0.18~$\mu$C/cm$^2$~\cite{Shirodkar14p157601}. Bruyer {\em et al.} later confirmed in theory that monolayers of transition-metal dichalcogenides $MX_2$ ($M$ = Mo, W; $X$ = S, Se, Te) in the $d1T$ phase are all ferroelectric, though only $d1T$-MoS$_2$ is lower in energy than its $c1T$ counterpart~\cite{Bruyer16p195402}. The ferroelectricity in 2D materials has also been confirmed in experiments. Notable examples are CuInP$_2$S$_6$~\cite{Belianinov15p3808}, $\alpha$-In$_2$Se$_3$~\cite{Zhou17p5508}, SnTe\cite{Chang16p274}, $d1T$-MoTe2~\cite{Yuan19p1775}, and WTe$_2$~\cite{Yang18p7160}. 

In practice, 2D FEs with out-of-plane (OP) polarization is generally favored over those with in-plane (IP) polarization for high-density integration via downscaling of the lateral dimensions. Using DFT calculations, Ding \etal predicted a family of III$_2$-VI$_3$ 2D materials such as $\alpha$-In$_2$Se$_3$ (Fig.~\ref{structures}) exhibiting both spontaneous OP and IP electric polarization at room temperature, which can be reversed with the assistance of a modest OP or IP electric field~\cite{Ding17p14956}.  Immediately following the prediction, Zhou \etal reported experimental evidences supporting the presence of OP polarization and ferroelectric domains in 10~nm multilayered $\alpha$-In$_2$Se$_3$ nanoflakes using piezoresponse force microscopy~\cite{Zhou17p5508}. Since then, a few more experimental studies have confirmed the existence of both OP and IP polarization in $\alpha$-In$_2$Se$_3$~\cite{Cui18p1253, Xue18p4976,Xiao18p227601,Poh18p6340,Xue18p1803738} with a thickness down to 3~nm~\cite{Xiao18p227601}. More recently, Wan \etal successfully fabricated a 2D ferroelectric FET consisted of graphene and layered $\alpha$-In$_2$Se$_3$, demonstrating nonvolatile memory after repeated writings of more than 10$^5$ cycles~\cite{Wan18p14885,Wan19p1808606}. 

Similar to their perovskite counterparts, 2D FEs may also find their applications in the fields of solar cells, photocatalysis, and optoelectronics.  An accurate prediction of their electronic structures is essential for guided design and development of 2D FE-based devices.
The DFT method within the Kohn-Sham formalism has become the first choice for reliable and efficient numerical simulations of condensed matter systems. The accuracy of DFT relies on the approximations to the exchange-correlation (XC) energy functional, and the local-density approximation (LDA)~\cite{Perdew92p13244} and the generalized-gradient approximation (GGA)~\cite{Perdew96p3865} are the most popular ones in the solid-state community. However, the remnant self-interaction error (SIE)~\cite{MoriSnchez06p201102} within LAD and GGA makes it challenging to correctly describe the electronic structures of systems with strongly localized electrons. A consequence of the SIE is the over-delocalization of $d$ and $f$ electrons~\cite{Perdew81p5048}, and thus a substantial underestimation of fundamental band gaps. By adding a fraction of the exact Hartree-Fock (HF) exchange, hybrid functionals reduce electron over-delocalization as the exchange term cancels the SIE originated from the Hartree term~\cite{Heyd03p8207,Paier06p154709,Marsman08p064201}. However, within the plane wave framework, hybrid functional calculations are much more demanding computationally than (semi-)local DFT due to the nonlocal nature of Fock exchange operator. Furthermore, the reliability of hybrid functionals in predicting band gaps of low-dimensional materials systems is questionable~\cite{Jain11p216806} considering that the rapid variation of screened Coulomb interactions is not captured by hybrid functionals assuming fixed dielectric screening. Specific to 2D FEs, the OP polarization will give rise to a built-in electric field that strongly bends the bands~\cite{Fu18p6312}, a feature may further complicating the electronic structure description. In this work, we aim to identify a cost-efficient first-principles method to accurately predict the electronic properties of 2D FEs. 

Here we focus on the DFT+$U$ method~\cite{Anisimov91p943, Liechtenstein95pR5467,Anisimov97p767,Dudarev98p1505, Kulik06p103001, Himmetoglu13p14}, an approach derived from the mean-field Hubbard model to correct the SIE at relatively low computational cost. The Hubbard parameter $U$ represents the strength of the on-site (screened) electron-electron Coulomb repulsion. By adding an energy penalty of $U$ for paring electrons in localized orbitals, the DFT+$U$ method alleviates the SIE of the chosen Hubbard manifold, leading to improved descriptions of some strongly correlated solids. Therefore, the value of $U$ is critical for the accuracy of the DFT+$U$ method. It is a common practice in literature to evaluate $U$ semi-empirically by fitting to experimental properties ({\em e.g.}, band gap) or results obtained with more accurate (albeit more expensive) first-principles methods such as hybrid functional calculations, and to assume $U$ being element-specific and transferable across different materials systems for simplicity. However, as intrinsic to the Hubbard model, the $U$ parameter should depend on the local atomic environment and is thus Hubbard-site specific. Since the Hubbard hamiltonian acts on atomic-like local orbitals that are often taken directly from pseudopotentials, the $U$ value is also sensitive to the construction of the pseudopotential ({\em e.g.}, the oxidation state of the reference state)~\cite{Kulik08p134314}. To make DFT+$U$ fully {\em ab initio}, it is highly desirable to determine $U$ self-consistently for a given atom in a given material from realistic electronic structure calculations. Several schemes for first-principles calculations of Hubbard parameters have been developed, such as linear response constrained density functional theory (LR-cDFT) approach~\cite{Cococcioni05p035105} and constrained random phase approximation~\cite{Springer98p4364, Kotani00p2413, Aryasetiawan04p195104, Aryasetiawan06p125106, Aichhorn09p085101,Takashi09p155134,Aichhorn10p064504,Sasioglu11p121101}.  Timrov \etal recently reformulated the LR-cDFT approach in the framework of density functional perturbation theory (DFPT), enabling efficient calculations of site-dependent Hubbard parameters for all inequivalent Hubbard sites without using a supercell~\cite{Timrov18p085127}. 

The newly developed Agapito-Curtarolo-Buongiorno-Nardelli (ACBN0) pseudohybrid Hubbard density functional allows a direct self-consistent evaluation of $U$ parameters~\cite{Agapito15p011006}. In ACBN0, the local Coulomb and exchange integrals are expressed in terms of  renormalized occupation matrices and renormalized occupations constructed from a localized basis set attached to the Hubbard atom. In a similar spirit, Lee \etal~\cite{Lee19p05967} and Tancogne-Dejean \etal~\cite{Tancogne-Dejean19p10813} respectively extended ACBN0 to allow a self-consistent determination of the inter-site interaction $V$ which represents the Coulomb repulsion strength between electrons on neighboring sites. Previous studies have shown that DFT+$U$+$V$ provides improved descriptions of electronic properties for materials such as charge-ordering and covalently bonded insulators~\cite{Anisimov96p4387,LeiriaCampoJr10p055602} where nonlocal correlations are important. For low-dimensional materials, the inclusion of $V$ helps to capture the effects due to local variations of screening of Coulomb interactions~\cite{Schuler13p036601,Wehling11p236805, Hansmann13p166401,Lee19p05967}. 

In this work, we perform a self-consistent DFT+$U$ study on 2D FEs represented by $\alpha$-In$_2$Se$_3$ and its homologous III$_2$-VI$_3$ compounds, aiming to identify an accurate and efficient first-principles method for these complex low-dimensional materials. Taking $\alpha$-In$_2$Se$_3$ as an example, we compare the $U$ values computed with LR-cDFT, DFPT, and ACBN0, and the resulting DFT+$U$ band structures. This serves as a comparative study of the accuracy of different schemes for Hubbard $U$ computations. It is found that the band gap is not sensitive to $U$ corrections applied to In 4$d$ or 5$p$ states, whereas the use of Hubbard $U$ on Se-$4p$ states greatly improves the PBE band value. The inclusion of inter-site interactions $V$ between valence $s$ and $p$ electrons of In and Se further increases the band gap. For all III$_2$-VI$_3$ compounds investigated, ACBN0 yield band gaps nearly matching HSE06 results. Notably, the inter-site $V$ correction is important for accurate descriptions of the electronic properties of van der Waals (vdW) heterostructures such as bilayer In$_2$Se$_3$ and In$_2$Se$_3$/InTe.  Finally, we demonstrate the importance of using self-consistent Hubbard parameters for quantitive predictions of adsorption energies of open-shell molecules on polar surfaces of 2D FEs with DFT+$U$. 

\section{Theory}
In this section,  we offer a brief introduction to the DFT+$U$ approach and selective first-principles methods for the calculations of Hubbard parameters. Interested readers should refer to the original papers for detailed discussions. 
\subsection{DFT+$U$}
The DFT+$U$~\cite{Anisimov91p943, Liechtenstein95pR5467,Anisimov97p767} method was formulated to correct the DFT energy by treating the electronic interactions between localized electrons in a separate way. The total energy is defined as
\begin{equation}
    \begin{aligned}
E_{{\rm DFT}+U} &= E_{\rm DFT} + E_{U} \\
&= E_{\rm DFT} + E_{\rm Hub} - E_{\rm dc}.
    \end{aligned}
\end{equation}
$E_{\rm DFT}$ is the standard DFT energy obtained with an approximate XC functional such as LDA or PBE. $E_{\rm Hub}$ is the Hubbard correction term that considers the on-site Coulomb interactions between electrons in localized orbitals of interest. $E_{\rm dc}$ is the double-counting correction that removes the electronic correlations already included in $E_{\rm DFT}$. In the rotational-invariant formulation proposed by Dudarev \etal~\cite{Dudarev98p1505}, the Hubbard energy $E_{\rm hub}$ as a function of occupation matrix $\textbf{n}^{I\sigma}$ of a localized basis set $\{ |\phi_m^I \rangle\}$ attached to atom $I$ can be written as
\begin{equation}
\label{Ehub}
    E_{\rm hub} = \frac{U^I}{2} \sum_{\sigma,m,m'} n_{mm}^{I\sigma} n_{m'm'}^{-I\sigma} + \frac{U^I-J^I}{2} \sum_{\sigma, m\neq m'} n_{mm}^{I\sigma} n_{m'm'}^{I\sigma}
\end{equation}
where $\sigma$ is the spin index, $m$ is the magnetic quantum number for a specific angular momentum $l$ ($-m \le l \le m$), and $U^I$ and $J^I$ are the spherically averaged on-site Coulomb repulsion and exchange interaction, respectively. For a periodic system, the occupation matrix $\textbf{n}^{I\sigma}$ is the projection of the occupied Kohn-Sham (KS) states $| \Psi_{\nu \textbf{k} }^{\sigma} \rangle$ to localized states $|\phi_m^I \rangle$:
\begin{equation}
\label{occupation}
    n_{m_1,m_2}^{I\sigma}=\sum_{\textbf{k}}^{N_\textbf{k}} \sum_\nu^{N_{\rm occ}} \langle \Psi_{\nu \textbf{k}}^{\sigma} | \phi_{m_2}^I \rangle \langle \phi_{m_1}^I | \Psi_{\nu \textbf{k}}^{\sigma} \rangle
\end{equation}
where $N_\textbf{k}$ is the total number of \textbf{k} points in the first Brillouin zone and $\nu$ is the band index that runs over $N_{\rm occ}$ occupied bands. The total number of localized electrons $N^{I\sigma}$ is expressed as $N^{I\sigma} = \sum_m n_{mm}^{I\sigma}$. The Hubbard manifold on which the Hubbard Hamiltonian acts is defined through the projector $\hat{P}_{m_1m_2}^I = | \phi_{m_2}^I \rangle \langle \phi_{m_1}^I |$. 

The double counting term is assumed as the Hubbard energy when the occupations of each localized orbital are either 1 or 0 ($(n_{mm}^{\sigma})^2 =n_{mm}^{\sigma} $), leading to
\begin{equation}    
    E_{\rm dc} = \frac{U^I}{2} N^I(N^I-1)-\frac{J^I}{2}\sum_{\sigma}N^{I\sigma}(N^{I\sigma}-1)
\end{equation}
with $N^I = \sum_{\sigma}N^{I\sigma}$. The Hubbard correction then has a simple form:
\begin{equation}
\label{DFTU}
    \begin{aligned}
        E_U &= E_{\rm Hub}-E_{\rm dc}\\
        &=  \sum_{I,\sigma}\frac{U^{I}_{\rm eff}}{2}\sum_{m, \sigma} \left \{ n_{mm}^{I\sigma} - \sum_{m'} n_{mm'}^{I\sigma} n_{m'm}^{I\sigma} \right \} \\
        &=\sum_{I,\sigma}\frac{U^{I}_{\rm eff}}{2} \mathrm{Tr}[ \textbf{n}^{I\sigma}(\textbf{1}-\textbf{n}^{I\sigma}) ]
    \end{aligned}
\end{equation}
where $U_{\rm eff}^I \equiv U^I-J^I$ is the effective Hubbard interaction parameter. To unclutter the narrations, we will refer $U_{\rm eff}$ as $U$ and omit the spin index $\sigma$ in the following discussions unless explicitly stated otherwise.  

\subsection{Hubbard $U$ from linear response theory}
The Hubbard correction depends strongly on the interaction parameter $U$. It is desirable to calculate $U$ in an internally consistent way.
Here we outline the key steps in LR-cDFT for first-principles calculations of Hubbard $U$~\cite{Cococcioni05p035105}. In LR-cDFT, $U$ is interpreted as the correction needed to recover the piece-wise linear behavior of the the total energy with respect to the orbital occupation, and thus can be computed from the second-order derivative of the energy. The total energy as a function of the localized orbital occupation $q^I$ of Hubbard site $I$ is given by 
\begin{equation}
\label{LR1}
    E(\{q^I\}) = \mathop{\min}_{\rho, \lambda^I} \left \{E_{\rm DFT}[\rho]+\sum_I \lambda^I (n^I - q^I)\right \}
\end{equation}
where $\rho$ is the (spin) charge density and  $\lambda^I$ is the Lagrange multiplier employed to constrain the site occupation $n^I$ defined as the trace sum of the occupation matrix (Eq.~\ref{occupation}). The physical interpretation of the Lagrange multiplier is to apply a perturbing potential of strength $\lambda^I$ to localized orbitals on site $I$. The Hubbard $U^I$ can then be computed as ${\partial^2 E}/{\partial (q^I)^2}$ via finite differences, a process requiring the evaluation of the total energy for a perturbed system with constrained $q^I$. In practice, it is more convenient to work with the Legendre transform of Eq.~\ref{LR1} which leads to a modified energy functional that depends on $\{\lambda^I\}$:
\begin{equation}
\label{LR2}
    E(\{\lambda^I\}) = \mathop{\min}_{\rho} \left \{E_{\rm DFT}[\rho]+\sum_I \lambda^I n^I \right \}
\end{equation}
Then the total energy as a function of on-site occupations $n^I$ (computed using the $\rho$ that minimizes Eq.~\ref{LR2}) is given via a Legendre transform,
\begin{equation}
    \bar{E}(\{n^I\})=E(\{\lambda^I\})-\sum_I \lambda^I n^I
\end{equation}
from which the first and second derivatives can be readily evaluated with 
\begin{equation}
    \frac{\partial \bar{E}}{\partial n^I} = -\lambda^I
\end{equation}
\begin{equation}
\label{ddE/ddn}
    \frac{\partial^2\bar{E}}{\partial(n^I)\partial(n^J)} = -\frac{\partial \lambda^I}{\partial n^J}.
\end{equation}
Technically, the perturbed ground state $|  \Psi_{\textbf{k}\nu} \rangle$  is obtained by solving following modified KS equation, 
\begin{equation}
\label{mKS}
    (\hat{H} + \lambda^J \hat{V}^J_{\rm pert} ) |  \Psi_{\textbf{k}\nu} \rangle = \varepsilon_{\nu\textbf{k}}  |  \Psi_{\textbf{k}\nu} \rangle 
\end{equation}
where $\hat{H}$ is the unperturbed KS Hamiltonian and $\hat{V}^J_{\rm pert}$ is the perturbing potential operator which is the sum of projectors on localized orbitals associated with atom $J$, $\hat{V}^J_{\rm pert} = \sum_m\hat{P}_{mm}^J$. All the values of $n^I$ are then collected for the perturbed ground state; by varying $\lambda^J$, we can construct the response matrix $\chi_{IJ} = \partial n^I/\partial \lambda^J$, and the Hubbard $U$ is the inverse of the response matrix: $U^I = - \chi_{II}^{-1}$, as derived from Eq.~\ref{ddE/ddn}. However, as pointed out by Cococcioni \etal, the complete definition of Hubbard $U$ is
\begin{equation}
    U^I=(\chi_0^{-1}-\chi^{-1})_{II}
\end{equation}
where $\chi_0$ is the response matrix of independent electron systems, resulting from the rehybridization of localized orbitals due to perturbations. In realistic DFT calculations, $\chi_0$ is evaluated with $n^I$ at the first iteration of the perturbed runs and $\chi$ is evaluated at self-consistency. It is noted that the off-diagonal elements $\chi_{IJ}^{-1}$ represent inter-site interactions $V$ in the extended Hubbard model~\cite{LeiriaCampoJr10p055602}. In practical calculations of periodic systems, a large supercell is often needed to make sure the localized perturbation not interacting with its images.

\subsection{Hubbard $U$ from density functional perturbation theory}
Recently, Timrov \etal reformulated LR-cDFT within the framework of DFPT~\cite{Timrov18p085127}. The main steps are (1) substitute finite differences with continuous derivatives; (2) recast perturbations in supercells as a sum of monochromatic (wave-vector-specific) perturbations in a primitive cell in reciprocal space.
The response matrix $\chi_{IJ}$ obtained by substituting Eq.~\ref{occupation} into Eq.~\ref{ddE/ddn} reads:
\begin{equation}
\label{response}
 \begin{aligned}
     \chi_{IJ}&=\sum_{m} \frac{\partial n_{m_1m_2}^{I}}{\partial \lambda^J} \\
     & =  \sum_{m} \sum_\textbf{k}^{N_\textbf{k}} \sum_\nu^{N_{occ}} 
   \left[ \langle \Psi_{\nu\textbf{k}} | \hat{P}_{m_2m_1} |\frac{\partial \Psi_{\nu\textbf{k}}}{\partial \lambda^J} \rangle \right.\\
    & \left. + \langle \frac{\partial \Psi_{\nu\textbf{k}}}{\partial \lambda^J} | \hat{P}_{m_2m_1} |  \Psi_{\nu\textbf{k}} \rangle \right]
 \end{aligned}
\end{equation}
where the LR KS wavefunction $| \frac{\partial \Psi_{\nu\textbf{k}}}{\partial \lambda^J} \rangle$ can be computed using the ordinary first-order perturbation approach (see details in ref.~\cite{Timrov18p085127}). This real-space implementation of $\chi_{IJ}$ within DFPT is essentially the same as that in LR-cDFT, leading to no computational advantage as localized perturbations still have to be applied to each unique Hubbard atom one at a time in the supercell. 

The second step is to recast perturbations in a supercell of size $L_1\times L_2 \times L_3$ as sums over monochromatic perturbations in a primitive unit cell on a grid of $\textbf{q}$ points defined by 
\begin{equation}
    \textbf{q}_{klm}=\frac{\bar{k}}{L_1}\textbf{b}_1 + \frac{\bar{l}}{L_3}\textbf{b}_2 + \frac{\bar{m}}{L_3}\textbf{b}_3
\end{equation}
where $\{\textbf{b}_1, \textbf{b}_2, \textbf{b}_3\}$  are the reciprocal lattice vectors of the primitive unit cell,  and $\bar{k}$, $\bar{l}$, $\bar{m}$ are integer numbers with $0 \leq \bar{k} < L_1, 0 \leq \bar{l }< L_2, 0 \leq \bar{m} < L_3$. The atomic index $I$ in a supercell now corresponds to two indices $(s,l)$ which means the $s$th atom in the $l$th primitive unit cell. Similarly, $J$ is replaced with $(s',l')$. The response matrix expressed in the reciprocal space has following form:
\begin{equation}
    \chi_{sl,s'l'}= \sum_m\frac{\partial n_{m_1m_2}^{sl}}{\partial \lambda^{s'l'}} = \sum_m\frac{1}{N_\textbf{q}}\sum_\textbf{q}^{N_\textbf{q}} e^{i\textbf{q}(\textbf{R}_l-\textbf{R}_{l'})} \Delta_\textbf{q}^{s'} \textbf{n}_{m_1m_2}^{s}
\end{equation}
where $N_\textbf{q}$ is the number of $\textbf{q}$ points, $\textbf{R}_l$ is the Bravais lattice vector of the $l$th primitive unit cell, and $\Delta_\textbf{q}^{s'} \textbf{n}_{m_1m_2}^{s}$ is the lattice-periodic response of the localized orbital occupation to a monochromatic perturbation of wave vector $\textbf{q}$. Detailed implementations of $\Delta_\textbf{q}^{s'} \textbf{n}_{m_1m_2}^{s}$ can be found in ref.~\cite{Timrov18p085127}. Compared to LR-cDFT, the DFPT approach is computationally cheaper by removing the need of supercell calculations, and has better numerical stability and convergence. 
\subsection{Hubbard $U$ from ACBN0}
Agapito \etal introduced an efficient approach to calculate the $U$ and $J$ values self-consistently~\cite{Agapito15p011006}. 
In ACBN0, electron-repulsion integrals are efficiently evaluated using pseudo-atomic orbitals (PAO) expressed as a linear combination of three Gaussian-type orbitals (3G). The KS orbitals are projected onto the PAO-3G basis set using a noniterative scheme by filtering out Bloch states with high-kinetic-energy components~\cite{Agapito13p165127}. This then allows the constructions of real-space density matrices and occupation numbers needed to compute the the Hartree-Fock energy associated with the chosen Hubbard manifold $\{m\}$ given by 
\begin{equation}
\label{EHF}
    \begin{aligned}
    E_{\rm HF}^{I\{m\}} &=\frac{1}{2}\sum_{\{m\}}\sum_{\alpha, \beta}
    \bar{P}_{mm'}^{I\alpha} \bar{P}_{m''m'''}^{I\beta}(mm'|m''m''')\\
    &-\frac{1}{2}\sum_{\{m\}}\sum_{\alpha}
    \bar{P}_{mm'}^{I\alpha} \bar{P}_{m''m'''}^{I\alpha}(mm'''|m''m')\\
    \end{aligned}
\end{equation}
where $\bar{P}_{mm'}^{I\sigma}$ ($\sigma = {\alpha, \beta}$) is the renormalized density matrices 
\begin{equation}
\label{nP}
 \bar{P}_{mm'}^{I\sigma} = \sum_{\textbf{k}}^{N_\textbf{k}} \sum_\nu^{N_{\rm occ}}\bar{N}_{\Psi_{\nu \textbf{k}}}^{I\sigma}
\langle \Psi_{\nu \textbf{k}}^{\sigma} | \phi_{m}^I \rangle \langle \phi_{m'}^I | \Psi_{\nu \textbf{k}}^{\sigma} \rangle
\end{equation} 
with $\bar{N}_{\Psi_{\nu \textbf{k}}}^{I\sigma}$ being the renormalized occupations 
\begin{equation}
\label{nN}
\bar{N}_{\Psi_{\nu \textbf{k}}}^{I\sigma} = \sum_{\{I\}}\sum_m\langle \Psi_{\nu \textbf{k}}^{\sigma} | \phi_{m}^I \rangle \langle \phi_{m}^I | \Psi_{\nu\textbf{k}}^{\sigma} \rangle.
\end{equation}
Following the treatment introduced by Mosey {\em et al.}~\cite{Mosey07p155123, Mosey08p014103}, the two sums in Eq.~\ref{nN} run over all atomic orbitals of the system that are attached to atoms of the same type as Hubbard site $I$ (referred to as $\{ \bar{m} \}$ in Ref.~\cite{Agapito13p165127}). The introduction of renormalized occupations into the density matrices accounts for the effects of screening. Note that $ 0 \le \bar{N}_{\Psi_{\nu \textbf{k}}}^{I\sigma} \le 1 $: in the limit of $\bar{N}_{\Psi_{\nu \textbf{k}}}^{I\sigma} = 1 $, Eq.~\ref{EHF} is the exact HF energy, whereas for $\bar{N}_{\Psi_{\nu \textbf{k}}}^{I\sigma} = 0 $, the DFT energy is recovered without Hubbard corrections as it should be for a fully delocalized Bloch state. The comparison of Eq.~\ref{EHF} and Eq.~\ref{DFTU} leads to density functionals of $U$ and $J$~\cite{Agapito13p165127,Tancogne-Dejean17p245133}:
\begin{equation}
    U=\frac{\sum_{\{m\}}\sum_{\alpha\beta}\bar{P}_{mm'}^\alpha \bar{P}_{m''m'''}^\beta (mm'|m''m''')}
    {\sum_{m \neq m'} \sum_\alpha n_{mm}^\alpha n_{m'm'}^\alpha +
    \sum_{\{m\}} \sum_\alpha n_{mm}^\alpha n_{m'm'}^{-\alpha} }
     \label{ACBN0U}
\end{equation}
\begin{equation}
     J=\frac{\sum_{\{m\}}\sum_\alpha\bar{P}_{mm'}^\alpha \bar{P}_{m''m'''}^\alpha( mm'''|m''m')}
    {\sum_{m \neq m'} \sum_\alpha n_{mm}^\alpha n_{m'm'}^\alpha}
     \label{ACBN0J}
\end{equation}
with the atomic index $I$ omitted. 

 \subsection{Hubbard $V$ from eACBN0}
 In the extended Hubbard model or DFT+$U$+$V$ approach, the inter-site Hubbard parameter $V$ represents the strength of Coulomb interactions between neighboring Hubbard sites. Note that in Eq.~\ref{Ehub}, the Hartree integrals $U^I$ is expressed as $U=\left\langle\phi_{m}\phi_{m^\prime}\left|V\right| \phi_{m}\phi_{m^\prime}\right\rangle$. Similarly, the inter-site interaction between atom $I$ and atom $J$ can be written as $V^{IJ}=\left\langle\phi_{i}^{I}\phi_{j}^{J}\left|V\right| \phi_{i}^{I} \phi_{j}^{J}\right\rangle$, where $i$ and $j$ are corresponding state indexes. The Hubbard energy in DFT+$U$+$V$ is derived as
\begin{equation}
\label{EhubUV}
E_{\mathrm{Hub}} =\sum_{I} \frac{U^{I}}{2}\left[\left(n^{I}\right)^{2}-\sum_{\sigma} \operatorname{Tr}\left[\left(\mathbf{n}^{I I \sigma}\right)^{2}\right]\right] \\
+\sum_{I,J} \frac{V^{I J}}{2}\left[n^{I} n^{J}-\sum_{\sigma} \mathrm{Tr}\left(\mathbf{n}^{I J \sigma} \mathbf{n}^{J I \sigma}\right)\right]
\end{equation}
The double-counting term including inter-site interactions is \begin{equation}
\label{DCUV}
E_{\mathrm{dc}}=\sum_{I} \frac{U^{I}}{2} n^{I}\left(n^{I}-1\right)+\sum_{I,J} \frac{V^{I J}}{2} n^{I} n^{J}
\end{equation}
Finally, we obtain the DFT+$U$+$V$ correction energy by substracting Eq.~\ref{DCUV} from Eq.~\ref{EhubUV} \cite{Campo10p055602}:
\begin{equation}\begin{aligned}
E_{U V} &=E_{\mathrm{Hub}}-E_{\mathrm{dc}}
&=\sum_{I, \sigma} \frac{U^{I}}{2} \mathrm{Tr}\left[\mathbf{n}^{I I \sigma}\left(\mathbf{1}-\mathbf{n}^{I I \sigma}\right)\right] -\sum_{I,J,\sigma} \frac{V^{I J}}{2} \mathrm{Tr}\left[\mathbf{n}^{IJ \sigma} \mathbf{n}^{J I \sigma}\right]
\end{aligned}\end{equation}
The inter-site Hubbard integral shares the {\it same} Coulomb interaction of $V$ with the on-site one but deals with its expectation value between orbitals at different positions unlike the on-site one for the same site. Thus, for the same orbitals belong to different positions, the values of $V^{IJ}$ should be proportional to $U^I$.

 Here we mostly follow the procedure of extended ACBN0 (eACBN0) developed by Lee \etal that enables self-consistent calculations of both $U$ and $V$~\cite{Lee19p05967}. In eACBN0, the renormalized occupation number for the pair of $I$ and $J$ is defined as
 \begin{equation}
 \bar{N}_{\Psi_{\nu \textbf{k}}}^{IJ\sigma} = \sum_{\{I, J\}}\sum_{i, j}\left [\langle \Psi_{\nu \textbf{k}}^{\sigma} | \phi_{i}^I \rangle \langle \phi_{i}^I | \Psi_{\nu \textbf{k}}^\sigma \rangle + \langle \Psi_{\nu \textbf{k}}^{\sigma} | \phi_{j}^J \rangle \langle \phi_{j}^J | \Psi_{\nu \textbf{k}}^\sigma \rangle \right ].
 \end{equation}
In close analogy to ACBN0, the renormalized density matrix for the pair is given by 
\begin{equation}
\bar{P}_{ij}^{IJ\sigma} = \sum_{\textbf{k}}^{N_\textbf{k}} \sum_\nu^{N_{\rm occ}}\bar{N}_{\Psi_{\nu \textbf{k}}}^{IJ\sigma}
\langle \Psi_{\nu \textbf{k}}^{\sigma} | \phi_{i}^I \rangle \langle \phi_{j}^J | \Psi_{\nu \textbf{k}}^{\sigma} \rangle.
 \end{equation}
After expressing the inter-site HF energy with renormalized density matrices and occupations for the pair, one can obtain a density functional for $V^{IJ}$,
\begin{equation}
    V^{IJ}=\frac{\sum_{ijkl}\sum_{\alpha\beta}\left [ \bar{P}_{ik}^{II\alpha} \bar{P}_{jl}^{JJ\beta}- \delta_{\alpha\beta} \bar{P}_{il}^{IJ\alpha} \bar{P}_{jk}^{JI\beta} \right ](ij|kl)}
    {\sum_{\alpha\beta}\sum_{ij} \left[ n_{ii}^{II\alpha}n_{jj}^{JJ\beta} - \delta_{\alpha\beta} n_{ij}^{IJ\alpha}n_{ji}^{JI\beta}\right] }
\label{Eq:VIJ}
\end{equation}
where $n_{ij}^{IJ\sigma}$ is the generalized occupation matrix defined as 
\begin{equation}
n_{ij}^{IJ\sigma} = \sum_{\textbf{k}}^{N_\textbf{k}} \sum_\nu^{N_{\rm occ}} \langle \Psi_{\nu \textbf{k}}^{\sigma} | \phi_{i}^I \rangle \langle \phi_{j}^J | \Psi_{\nu \textbf{k}}^{\sigma} \rangle
\end{equation}
For eACBN0 computations, we use the onsite Hubbard interaction as $U_\textrm{eff}\equiv U-J$ where $U$ and $J$ are given by Eq.~\ref{ACBN0U} and Eq.~\ref{ACBN0J}, respectively, and the inter-site one by Eq.~\ref{Eq:VIJ}.

\section{Computational Methods}
All DFT calculations are performed with \texttt{QUANTUM ESPRESSO}~\cite{Giannozzi09p395502, Giannozzi17p465901} using PBE XC functional~\cite{Perdew96p3865} and norm-conserving (NC) pseudopotentials. A slab model with a vacuum layer along the $z$ axis of at least 15~\AA~is used to model 2D FEs. The dipole correction in the center of the vacuum is employed to remove the artificial electric field and unphysical dipole-dipole interactions between the slab and its periodic images. The in-plain lattice constants and atomic positions are fully relaxed using PBE XC functional with an energy convergence threshold of $10^{-5}$~Ry, a force convergence threshold of $10^{-4}$ Ry/Bohr, and a $8\times8\times1$ Monkhorst-Pack $k$-point grid for Brillouin zone sampling. The kinetic energy cutoff for plane wave expansion is 80 Ry. The cold smearing of Marzari-Vanderbilt is chosen as the orbital occupation scheme with a temperature equal to 1 mRy/$k_b$. All electronic properties (\eg~band structure) are then computed using PBE optimized structures. We find that the dipole correction has little impacts on the optimized structures but generally reduces the band gap slightly. 

For hybrid functional HSE06~\cite{Krukau06p224106} band gap calculations, we use a $4\times4\times1$ $q$-point grid in combined with a $8\times8\times1$ $k$-point grid. The HSE06 band structure is obtained via Wannier interpolation~\cite{Marzari12p1419} using \texttt{Wannier90}~\cite{Pizzi20p165902} interfaced code with \texttt{Quantum ESPRESSO}. The self-consistent $U$ values within the ACBN0 approach are computed using \texttt{AFLOW$\pi$}~\cite{Supka17p76} and the NC pseudopotential library coming with the package (the same ones for structural optimizations). In the case of DFPT for Hubbard parameters, the L\"{o}wdin orthogonalized atomic wave functions are used in the self-consistent field (SCF) calculations, followed by perturbative calculations using a $2\times2\times1$ mesh for $\mathbf{q}$ space sampling and a convergence threshold of 1.0$\times 10^{-8}$ for the response function. The Hubbard $V$ parameters are calculated with eACBN0 using an in-house version of \texttt{QUANTUM ESPRESSO}~\cite{Lee19p05967} and GBRV ultrasoft pseudopotentials~\cite{Garrity14p446}. Fully converged values are obtained when the difference between the energies in the self-consistent loop is less than 10$^{-8}$ Ry. We consider all the inter-site interactions of Eq.~(\ref{Eq:VIJ}) between the $I$ and $J$ atoms of which inter-atomic distance is less than 6.0 \AA~and set the onsite $U$ for $s$-orbitals to be zero as discussed before~\cite{Lee19p05967}. 

\section{Results and Discussions}
\subsection{Benchmark study of $\alpha$-In$_2$Se$_3$}

\begin{figure}[ht]
\centering
\includegraphics[scale=1.0]{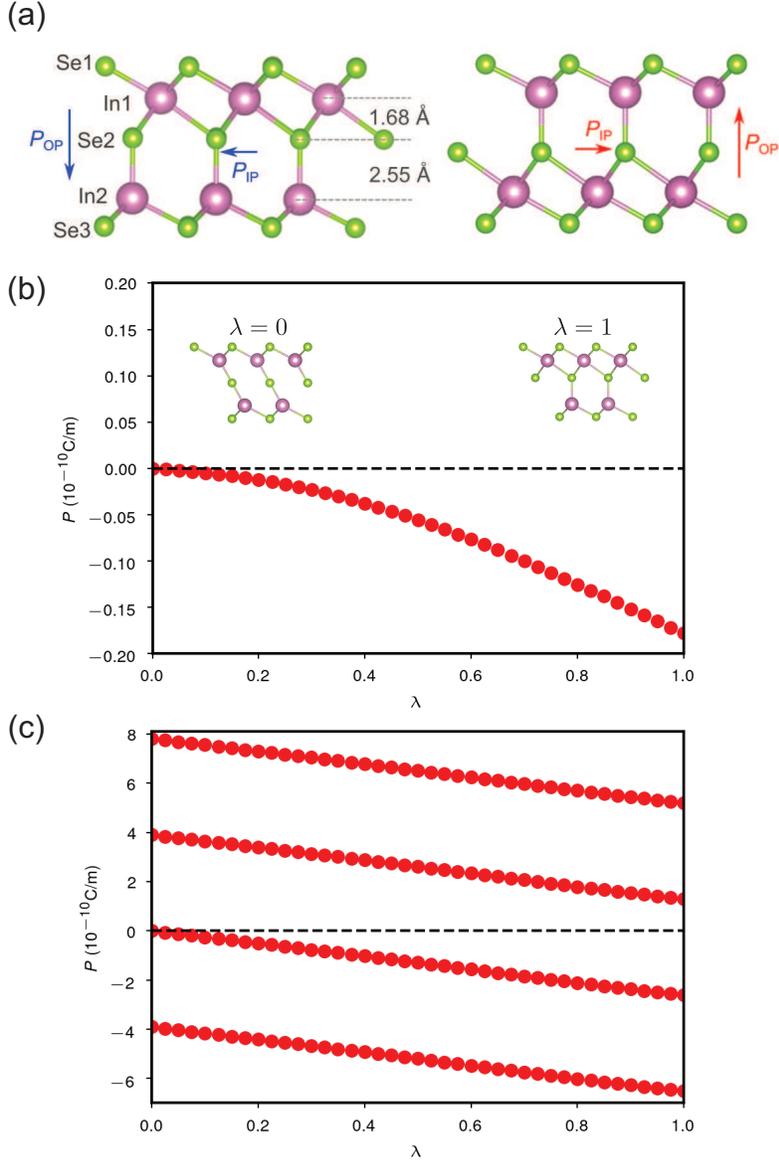}
 \caption{Schematic of two-dimensional ferroelectric $\alpha$-In$_2$Se$_3$.  Each atomic layer in the quintuple layer has only one type of atom arranged in a triangular lattice. The displacement of the central Se layer gives rise to both in-plane polarization ($P_{\rm IP}$) and out-of-plane polarization ($P_{\rm OP}$). The switch of $P_{\rm OP}$ will lead to the reversal of $P_{\rm IP}$.  Calculation of the (b) out-of-plane and (c) in-plane effective polarization with the Berry-phase approach. The structure is changed adiabatically from a non-polar phase ($\lambda=0$) to a polar phase ($\lambda=1$).
}
  \label{structures}
 \end{figure}

We first focus on the electronic structure of 2D $\alpha$-In$_2$Se$_3$ given that its room-temperature ferroelectricty has been confirmed experimentally. The 2D $\alpha$-In$_2$Se$_3$ is a covalently-bonded quintuple layer stacked in the sequence of Se-In-Se-In-Se with each layer containing only one type of atom arranged in a triangular lattice. Our optimized IP lattice constant with PBE is 4.10 \AA, and the band gap is 0.80~eV, in agreement with previous DFT results~\cite{Ding17p14956,Tiwari20p235448}. The structural origin of ferroelectricity in $\alpha$-In$_2$Se$_3$ lies at the displacement of the central Se layer with respect to the top and bottom In-Se layers, which breaks the centrosymmetry along both OP and IP directions (Fig.~\ref{structures}). 
We calculate the polarization with the Berry-phase approach by
tracking the change in Berry phase during a structural transformation process in which a non-polar structure is changed adiabatically to a polar structure (Fig.~\ref{structures}b-c). The 2D polarization is defined as $P = D/S_{\rm IP}$, where $D$ is the dipole moment and $S_{\rm IP}$ is the in-plane area of a 2D unit cell. It is noted that direction of in-plane polarization was controversial in previous studies~\cite{Ding17p14956,Tiwari20p235448,Liu19p025001}. Our berry-phase calculations unambiguously determine the direction of $P_{\rm IP}$ as shown in Fig.~\ref{structures}a. The values of $P_{\rm IP}$  and $P_{\rm OP}$  are 0.26 nC/m and 0.018 nC/m, respectively.

\begin{figure}[ht]
\centering
\includegraphics[scale=1.0]{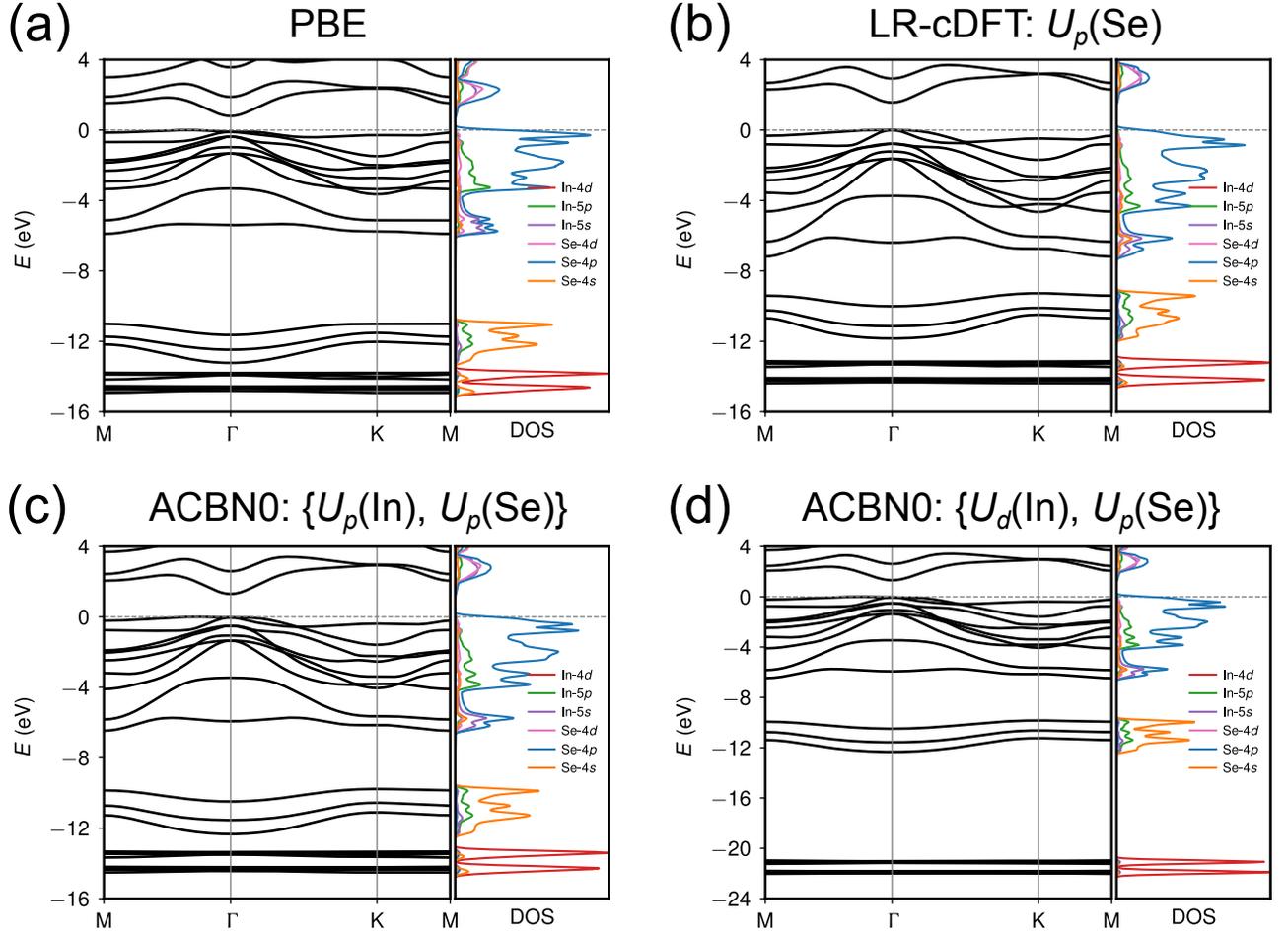}
 \caption{Comparison between the band structures and projected density of states of $\alpha$-In$_2$Se$_3$ obtained with (a) PBE (b) Hubbard $U$ computed with LR-cDT and applied to Se-$4p$ states (c) ACBN0 with $U$ corrections applied to  Se-$4p$ and In-$5p$ states (d) ACBN0 with $U$ corrections applied to Se-$4p$ and In-$4d$ states. The density of states of In-$4d$ are scaled to 20\% of their original values in the plots.}
 \label{compareBandsDOS}
 \end{figure}

The projected density of states (PDOS) obtained with PBE (Fig.~\ref{compareBandsDOS}a) reveals that the valence states near the Fermi level ($E_F$) are predominantly consisted of In-$5p$ and Se-$4p$ states and the conduction states are mostly of Se-$4d$, Se-$4p$, and In-$5s$ characters. The low-dispersion deep levels between $-16$ to $-10$ eV are from Se-4$s$ and semi-core In-$4d$ states. We note here that such electronic structure is very different from those of conventional ferroelectrics such as PbTiO$_3$ where the states near $E_F$ are of Ti-3$d$ and O-2$p$ characters. It is well established that the emergence of ferroelectricity in perovskites is due to a delicate balance between the long-range Coulomb interactions (favoring the break of inversion symmetry) and the short-range repulsions (favoring the non-polar high-symmetry phase), and the $p$-$d$ hybridization between transition metal and oxygen atoms weakens the short-range repulsions thus responsible for the ferroelectric distortions~\cite{Cohen92p136}. It is evident that $\alpha$-In$_2$Se$_3$ with mostly $p$-$p$ hybridization does not fit into this picture, hinting at a different mechanism for ferroelectricity at reduced dimensions. 

The DFT+$U$ method has the flexibility to choose the Hubbard manifold, namely the localized orbitals on which the Hubbard Hamiltonian will act. 
It is more common to apply $U$ corrections to open $d$ or $f$-electron shells. However, previous investigations with ACBN0 demonstrated the importance of introducing on-site repulsions to electrons on $p$-states and closed-shell $d$-states (\eg~O-2$p$ and Zn-3$d$ in ZnO~\cite{Agapito15p011006}) as well.
To serve as a complete test, we investigate following two cases: $U$ applied to Se-$4p$ and In-$4d$ states denoted as $\{U_p({\rm Se}), U_d({\rm In}) \}$, and $U$ applied to Se-$4p$ and In-$5p$ states denoted as $\{U_p({\rm Se}), U_p({\rm In}) \}$. The $U$ parameters are computed using LR-cRT, DFPT, ACBN0, and eACBN0, respectively. It is noted that the break of inversion symmetry along the OP direction makes all five atoms unique Hubbard sites. 

Table~\ref{Uvalues} reports the $U$ parameters estimated with different first-principles schemes and the corresponding band gaps ($E_g$). The LR-cRT method has been widely used in the literature because of its computational simplicity, however, its application to fully occupied localized orbitals is more problematic due to the small linear response to perturbations~\cite{Lee12p781}. This is the case for In-$4d$ which has a shell filling of $d^{10}$: the single-shot LR-cRT calculations using a unit cell and a $2\times2\times1$ supercell both yield unphysically large $U$ values (thus omitted in Table~\ref{Uvalues}). In comparison, DFPT and ACBN0 are capable of determining $U_d$(In) with comparable magnitudes. The high values of $U_d$(In) ($\approx13$--$15$~eV) are unsurprising given that the magnitude of $U$ is proportional to the occupancy of the localized orbital; large $U$ values were previously reported for $d^{10}$ transition metals such as Zn$^{2+}$ ($U_d=12.8$~eV) in ZnO~\cite{Agapito15p011006}. In the case of $U_p$(Se), the values obtained with the LR method depend on the size of supercell, i.e., $U_p$(Se1) changes from 7.83~eV in a unit cell to 5.32~eV in a $2\times2\times1$ supercell. The LR values are significantly higher than those obtained with self-consistent schemes ($U_p=3$--$4$~eV). In a recent study of LiFePO$_4$, it was also observed that the single-short LR value of $U$ for Fe-3$d$ is $\approx7$~eV, much higher than the DFPT value of $\approx5$~eV~\cite{Cococcioni19p033801}.  
The Hubbard $U$ parameters for In-$5p$ states estimated with different methods are comparable, and they are all small values which are expected from the $5p^0$ configuration of In$^{3+}$ if In$_2$Se$_3$ is fully ionic. 
It is noted that different Se (In) atoms acquire different values of $U_p$ ($U_d$) despite being of the same species in the same material, which is the consequence of the OP polarization breaking inversion symmetry. This confirms Hubbard interactions are sensitive to the fine details of local atomic environments, and should not be considered as transferable/tunable parameters. 

\begin{figure}[ht]
\centering
\includegraphics[scale=1.5]{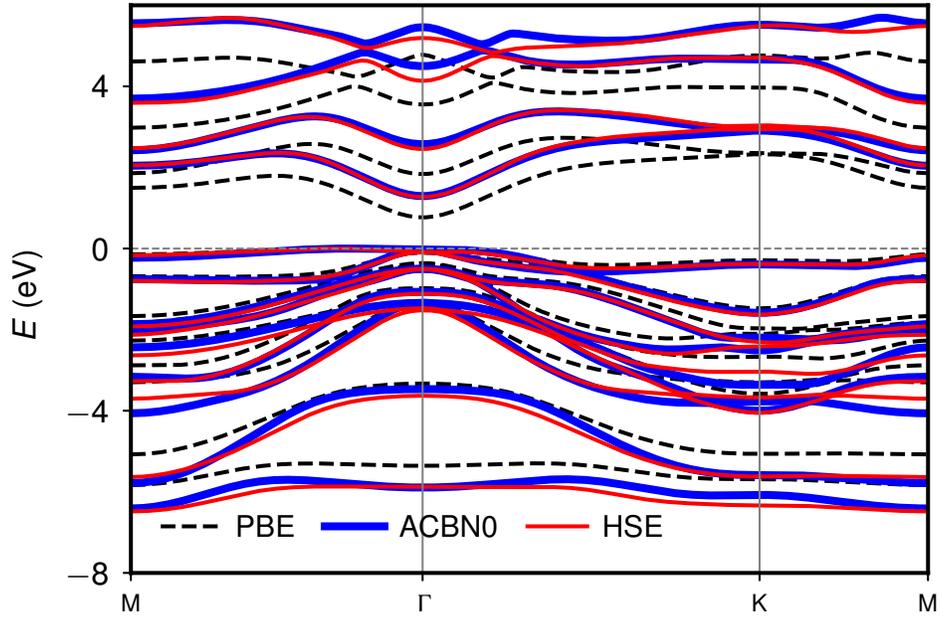}
 \caption{Comparison of the band structures of $\alpha$-In$_2$Se$_3$ computed with PBE, ACBN0, and HSE06, respectively. The valence band maximum is set as the Fermi level.}
  \label{compareBands}
 \end{figure}

 The band gaps computed using self-consistent Hubbard parameters ($E_g=$~1.34 and 1.32 eV for DFPT and ACBN0, respectively) all improve upon the PBE value (0.80 eV), agreeing well with the HSE06 result (1.48 eV).  We find that the band gap of $\alpha$-In$_2$Se$_3$ is insensitive to Hubbard corrections applied to In-$4d$ states but correlates positively with $U_p$(Se). As shown in Fig.~\ref{compareBandsDOS}d, the effect of a high value of $U_d$(In) is to downshift the deep-lying flat $4d$ bands that are already disentangled from the $4p$ states of Se near the Fermi level at the PBE level (Fig.~\ref{compareBandsDOS}a). Therefore, the band gap does not dependent on $U_d$(In). The improvement of band gap prediction with ACBN0 mainly comes from the on-site Coulomb interactions on Se. This is different from ZnO where a large $U$ correction to Zn-$3d$ is needed to reduce the $p$-$d$ repulsion between the low-lying Zn-$3d$ bands and the O-$2p$ bands that dominate the valance band edge~\cite{Agapito15p011006}. The high value of $U_p$(Se) from LR-cDFT leads to a larger band gap. The comparison of the band structures of PBE, ACBN0, and HSE06 is shown in Fig.~\ref{compareBands}. We find that ACBN0 has almost the same band dispersion as the hybrid functional HSE06 for states over a broad energy range from $-8$ to $4$~eV. 

\begin{figure}[ht]
\centering
\includegraphics[scale=0.8]{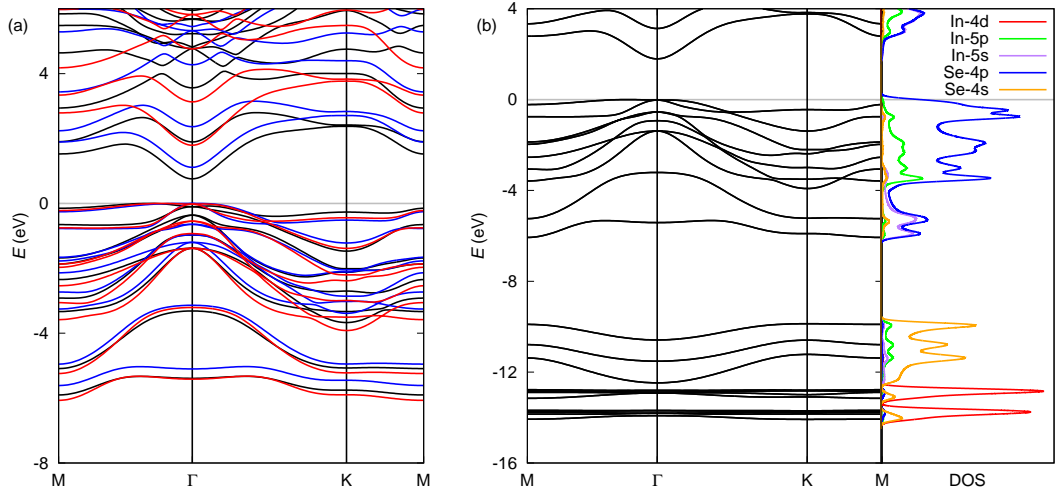}
 \caption{(a) Comparison of the band structures of $\alpha$-In$_2$Se$_3$ computed with PBE (black), ACBN0 (blue), and eACBN0 (red), respectively. The valence band maximum is set as the Fermi level. (b) Projected density of states computed from eACBN0. The density of states of In-$4d$ are scaled to 20\% of their original values in the plots.}
  \label{eACBN0Bands}
 \end{figure}

We further perform DFT+$U$+$V$ computations using the eACBN0 method 
to study the role of inter-site interactions $V$ in improving the band gap
of $\alpha$-In$_2$Se$_3$.
We obtain a band gap of 1.79 eV that is larger than those from DFT+$U$ and HSE06. 
This trend is consistent with a recent study on low-dimensional materials (\eg~2D black phosphorous)~\cite{Lee19p05967}.
Considering that HSE06 may not capture the rapid variation of Coulomb screening in low-dimensional materials~\cite{Jain11p216806}, the inter-site Hubbard interaction in eACBN0 approach could compute the band gap more accurately. 
In Table~\ref{UVvalues}, the calculated Hubbard $U$ and $V$ values are summarized.
Since the band gap is insensitive to the inclusion of In $4d$-orbital, we {\color{blue} choose} $p$-orbitals as a mainfold for onsite interaction and $s$- and $p$-orbitals for inter-site interactions, respectively.
As shown in Table~\ref{UVvalues}, the obtained on-site $U$ values are 
quite comparable to the values from DFPT and ACBN0 in Table~\ref{Uvalues}. 
Moreover, the converged inter-site $V$ well reflects the variation of local screening in direction perpendicular to the plane. 
We also confirm that the converged $V$ decreases rapidly as the interatomic distance increases (not listed in Table~\ref{UVvalues}).
In Fig.~\ref{eACBN0Bands}a, we compare the band structures obtained with PBE, ACBN0 and eACBN0, respectively. 
Like bands from ACBN0 and HSE06 in Fig.~\ref{compareBands}, 
the conduction bands obtained from the eACBN0 method are almost rigidly shifted up 
with respect to those from PBE approximations. 
We also compute the PDOS from the eACBN0 method in Fig.~\ref{eACBN0Bands}b and find that the contributions from each orbitals are similar to those from ACBN0 in Fig.~\ref{compareBandsDOS}c while all conduction bands are rigidly shifted. 
We note that the inclusion of In-$4d$ for $U$ and $V$ interactions does not affect the band dispersions and PDOS, though giving a slightly larger band gap of 1.86~eV. 

From this benchmark study, we make following arguments. The on-site Hubbard $U$ can be considered as a ``fingerprint" of local atomic environment: it does not have a simple dependence on the element type or crystal structure, which highlights the necessity of using self-consistent values to be fully {\em ab initio}. The three self-consistent schemes for computing Hubbard parameters, DFPT, ACBN0, and eACBN0, give comparable values, and are numerically more stable than the single-shot LR method. The flexibility to choose the Hubbard manifold to some extent increases the variational degrees of freedom of DFT+$U$, though in practice it may be less straightforward to choose the optimal sets of localized states to apply Hubbard corrections. We argue band gaps computed using self-consistent $U$ are less sensitive to the choice of Hubbard manifold, as supported by Table~\ref{Uvalues} where two choices of Hubbard manifolds result in similar band gap values. 

\subsection{Band gaps of III$_2$-VI$_3$ 2D FEs}
In the seminal work of Ding {\em et al.}, a family of III$_2$-VI$_3$ compounds, Al$_2$S$_3$, Al$_2$Se$_3$, Al$_2$Te$_3$, Ga$_2$S$_3$, Ga$_2$Se$_3$,Ga$_2$Te$_3$, In$_2$S$_3$, and In$_2$Te$_3$, were predicted to be ferroelectric when they adopt the same structure as that of $\alpha$-In$_2$Se$_3$.  This offers a platform to test the performance of DFT+$U$ using self-consistent Hubbard parameters. We calculate the band gaps for all known III$_2$-VI$_3$ 2D FEs using PBE and ACBN0, respectively. In our benchmark study of  $\alpha$-In$_2$Se$_3$, each atom is treated as a unique Hubbard site due to symmetry breaking. Nevertheless, we expect in some cases it might be computationally tedious to treat all symmetry-inequivalent sites as unique Hubbard sites. At the lowest approximation, it is still reasonable to use the same $U$ for atoms of the same species, which is equivalent to averaging on-site interactions over these atoms. To gauge the subtle effect of applying averaged $U$ corrections, we compare the results obtained using two (element-specific) $U$'s and five (site-specific) $U$'s, denoted as ACBN0+2$U$ and ACBN0+5$U$, respectively. The HSE06 value is chosen as the reference because of the lack of experimental band gaps for these newly discovered 2D FEs. 

\begin{figure}[ht]
\centering
\includegraphics[scale=1.5]{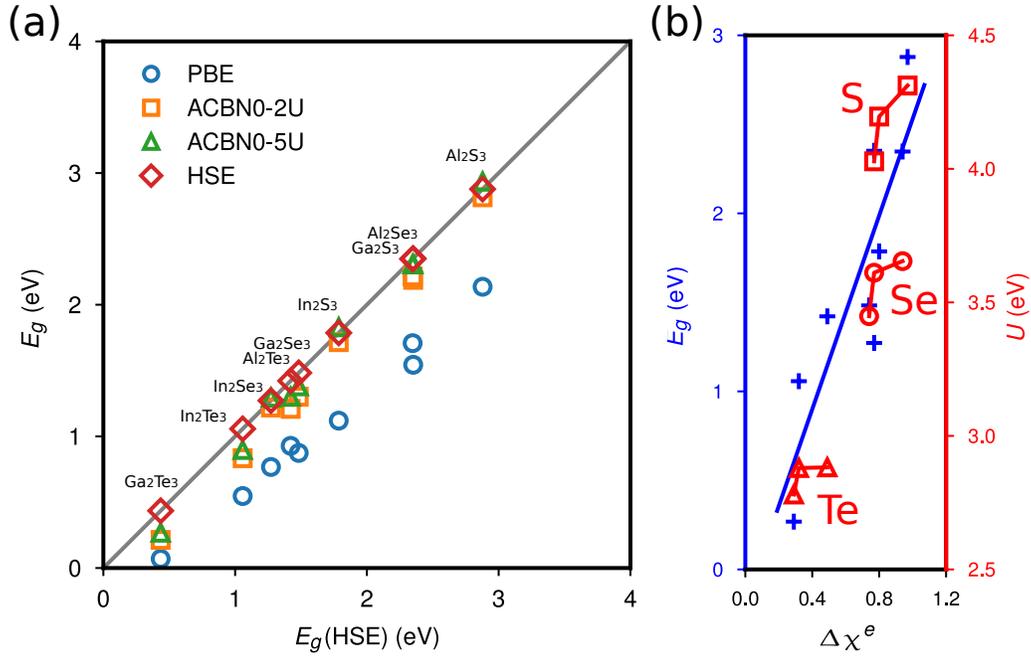}
 \caption{(a) Comparison of band gaps predicted by PBE, ACBN0+2$U$, and ACBN0+5$U$ using HSE06 values as the reference. (b) Correlation between band gaps, Hubbard $U$ values of group-VI elements computed with ACBN0, and electronegativity difference ($\Delta \chi^{e}$) of III$_2$-VI$_3$ compounds. The solid line in blue is the linear fit of $E_g$ versus $\Delta \chi^{e}$.}
  \label{compareGap}
 \end{figure}

Figure.~\ref{compareGap}a compares the band gaps calculated using PBE, ACBN0+2$U$, ACBN0+5$U$, and HSE06. It is clear that PBE substantially underestimates the band gaps for all III$_2$-VI$_3$ compounds. The band gaps predict with  ACBN0+2$U$ and ACBN0+5$U$ are comparable to HSE06 results, with ACBN0+5$U$ being slightly better that ACBN0+2$U$. This confirms the applicability of ACBN0 to a broader range of III$_2$-VI$_3$-type 2D FEs. We also note the correlation between the band gap, the electronegative ($\chi^{e}$), and the self-consistent value of Hubbard $U$ of group-VI elements (S, Se, and Te). First, the band gap increases with the difference in the electronegative ($\Delta \chi^{e}$) of group-III and group-VI elements (Fig.~\ref{compareGap}b). Taking Ga$_2$Te$_3$ as an example, it has the smallest band gap, and $\Delta \chi^{e} = 0.29$ ($\chi^{e}(\rm Ga)$ = 1.81, $\chi^{e}(\rm Te)$ = 2.1) is also the smallest among all 9 compounds. In comparison, Al$_2$S$_3$ has the largest $\Delta \chi^{e}$ as well as the largest band gap. Second, the $U$ value of an element depends on its electronegativity, as we find $U_p$(Te) $<U_p$(Se) $<U_p$(S) that is consistent with the trend of $\chi^{e}({\rm Te}) < \chi^{e}({\rm Se}) < \chi^{e}({\rm S})$. For the same group-VI element, the $U$ value also scales with $\Delta \chi^{e}$ of the material. For example, $U_p$(S) in In$_2$Se$_3$ is smaller than that in Al$_2$S$_3$ while the former has a smaller $\Delta \chi^{e}$. This further supports our previous argument that $U$ is not element but site/material specific. We may also take advantage of this feature to use the self-consistent $U$ to differentiate atoms of the same element in complex materials such as charge-ordering transition metal oxides. 

\subsection{VdW heterostructures of 2D FEs}
The 2D vdW heterostructures consisted of multiple layers of 2D materials are becoming an increasingly important platform to realize novel emergent phenomena~\cite{Duong17p11803}. From a device perspective, the absence of surface dangling bonds in 2D materials and the weak vdW interactions between different layers ensure an atomically sharp interface across the heterojunction, beneficial for obtaining stable and reproducible device characteristics~\cite{Chhowalla16p16052}. The incorporation of 2D FEs into vdW heterostructures allows for convenient control of electrical properties such as band gap, band alignments, and charge transport via external electric fields. An important consequence of an OP polarization is a built-in electric field and thus an electrostatic potential across the material. It was suggested that one can take advantage of the vacuum level difference between the two surfaces of 2D FEs to realize water splitting with near-infrared light~\cite{Li14p018301}. Our investigations demonstrate that ACBN0 improves the band gap prediction for freestanding monolayer III$_2$-VI$_3$ 2D FEs. Using $\alpha$-In$_2$Se$_3$ bilayer and InTe/In$_2$Se$_3$ as examples, we further analyze the performance of PBE, ACBN0, and eACBN0 on vdW heterostructures. The in-plane lattice constants and atomic positions are optimized using PBE with the inclusion of Grimme dispersion corrections (DFT-D3)~\cite{Grimme06p1787}. It is noted that the dipole correction has negligible impacts on the optimized geometries but slightly reduces the electronic band gaps.

\begin{figure}[ht]
\centering
\includegraphics[scale=1.0]{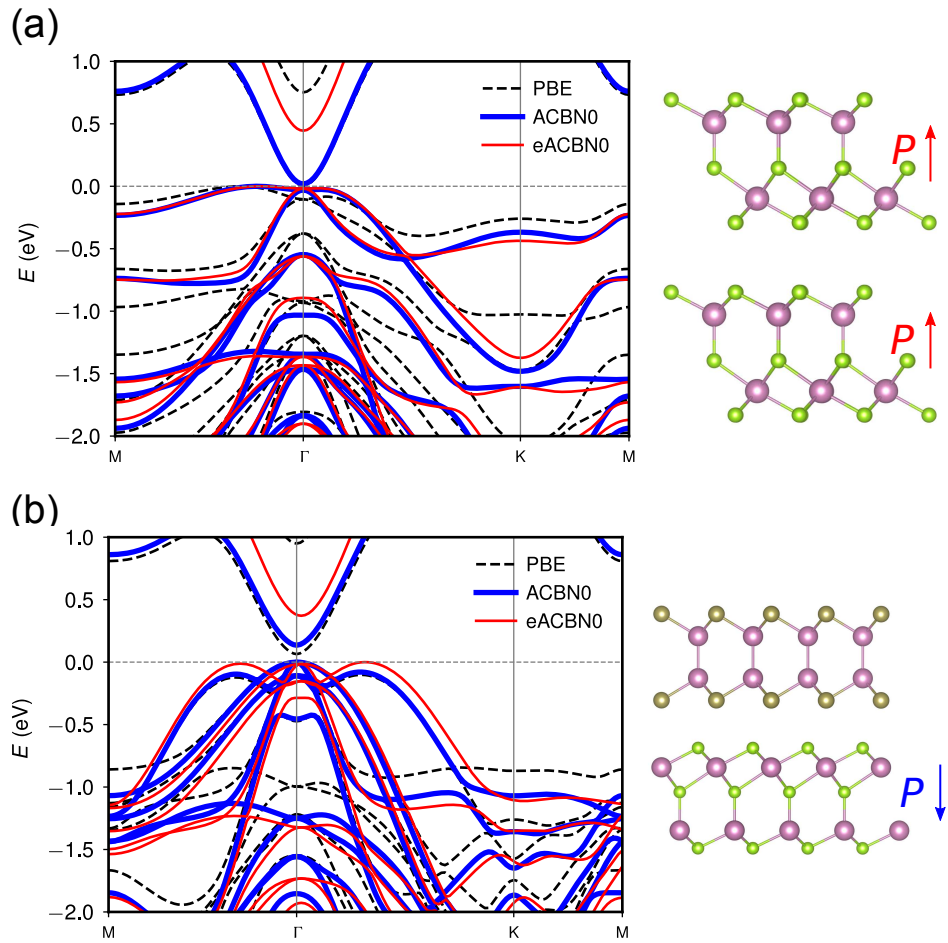}
 \caption{Comparison of PBE, ACBN0, and eACBN0 band structures of (a) $\alpha$-In$_2$Se$_3$ bilayer and (b) In$_2$Se$_3$/InTe vdW heterstructure. }
  \label{compareVDW}
 \end{figure}

For $\alpha$-In$_2$Se$_3$ bilayer with the same polarization direction for each layer, PBE predicts a nearly semimetal state ($E_g<5$~meV) while ACBN0 yields a small (indirect) band gap of $\approx$0.02 eV, as shown in Fig.~\ref{compareVDW}a.  The direct band gap at the $\Gamma$ point predicted by PBE and ACBN0 are 0.11 and 0.05 eV, respectively, both lower than the HSE06 value of 0.26~eV (obtained with {\color{blue}$8\times8\times1$} $k$/$q$-point grids). This suggests both methods fail to capture the correct charge transfer between outermost layers due to the strong perpendicular electric field. In comparison, eACBN0 predicts a band gap of 0.44~eV, indicating a better description of the charge distribution along the OP direction. This is supported by the close agreement between eACBN0 (1.13 D) and HSE06 (0.95 D) values of the calculated dipole moments. Likewise, the computed band gaps for $\alpha$-In$_2$Se$_3$/InTe heterostructure highlight the important role of inter-site Hubbard interactions in 2D FE materials. The band gaps of the system from PBE and ACBN0 are found to be 0.06 and 0.14 eV, respectively, while the gap from eACBN0 is 0.37~eV (Fig.~\ref{compareVDW}b), comparable to the HSE06 result of 0.29~eV. Here, we also observe eACBN0 predicts a larger band gap than HSE06, consistent with previous studies.

 \subsection{Adsorption energy of small molecules}
It is well established that ferroelectricity can affect the electronic structure of the surface and thus adsorption energies and catalytic properties, making ferroelectrics promising candidates for tunable catalysis~\cite{Levchenko08p256101,Kolpak08p036102,Kakekhani16p302,Kakekhani16p19676}. It was recently proposed that $\alpha$-In$_2$Se$_3$ may act as a reversible gas sensing substrate because NH$_3$, NO and NO$_2$ have different adsorption energies on the two oppositely charged surfaces~\cite{Tang20p7331}. The ability to accurate predicate the adsorption energy is essential for the understanding of the adsorption mechanisms on catalytic surfaces. However, local and semi-local density functionals often failed quantitatively to predict the adsorption energies of small molecules on metal surfaces  (\eg CO on Cu) because of the incorrect predictions of the positions of the frontier orbitals of molecules~\cite{Mason04p161401}. 

\begin{figure}[ht]
\centering
\includegraphics[scale=2.0]{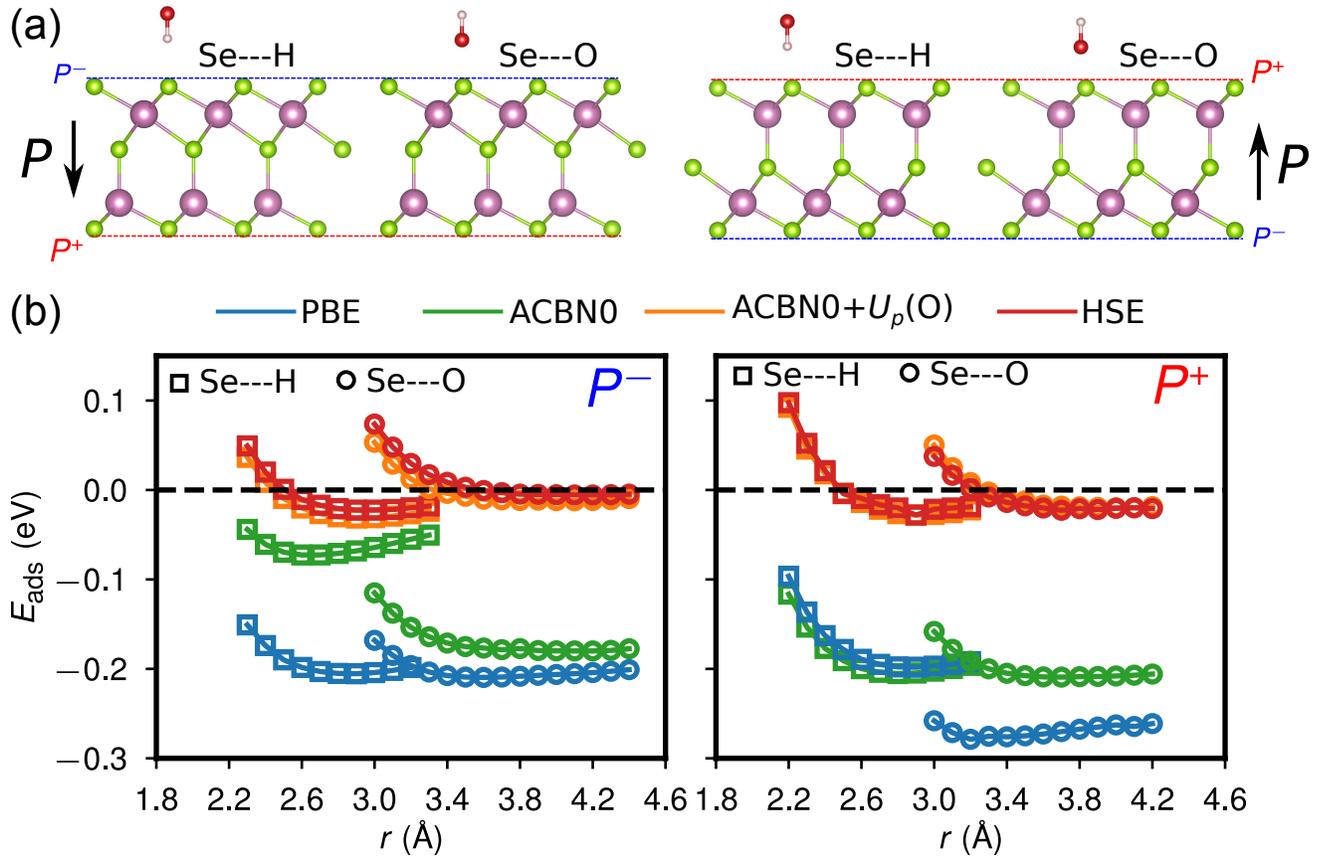}
 \caption{(a) Schematics of OH adsorptions on the polar surfaces of $\alpha$-In$_2$Se$_3$. (b) Adsorption energy ($E_{\rm ads}$) versus adsorption distance employing PBE, ACBN0, ACBN0+$U_p$(O), and HSE06.}
  \label{adsorption}
 \end{figure}

Here we first carry out a benchmark study of the adsorption of hydroxyl radical (HO) on both surfaces of $\alpha$-In$_2$Se$_3$ by constructing the energy profile as a function of adsorption distance using PBE, ACBN0, and HSE06, respectively. The adsorption energy is defined as $E_{\rm ads} = E_{\rm total} - E_{\rm molecule} - E_{\rm In_2Se_3}$, where $E_{\rm total}$, $E_{\rm molecule}$, and $E_{\rm In_2Se_3}$ are the energies of the gas molecules adsorbed on the In$_2$Se$_3$ monolayer, isolated gas molecules, and In$_2$Se$_3$ monolayer, respectively. Notably, we also investigate the effects of adding self-consistent $U$ corrections to the oxygen 2$p$ states of HO, denoted as ACBN0+$U_p({\rm O})$. Four adsorption cases (Fig.~\ref{adsorption}a) are explored, O-end adsorptions on $P^+$ and $P^-$ surfaces and H-end adsorptions on $P^+$ and $P^-$ surfaces. The HO remains perpendicular to the slab when varying the adsorption distance. The results are presented in Fig.~\ref{adsorption}b. We find that PBE strongly overestimates not only the binding strength compared to HSE06 but also the change in adsorption energy of HO due to polarization switching:  the PBE value of $E_{\rm ads}$ for O-end adsorption changes from $-0.21$ eV on $P^-$ surface to $-0.28$ eV on $P^+$ surface, though the HSE value changes slightly from $-0.006$ eV to $-0.02$ eV.  ACBN0 calculations with self-consistent $U_p({\rm Se})$ and $U_d({\rm In})$ improve the overall predictions of adsorption energies compared to PBE though the results remain qualitatively different from HSE06.  It turns out the ACBN0 $U_p$ value of the O-2$p$ state in an isolated HO molecule is as large as 7.74~eV. After applying Hubbard $U$ corrections to the oxygen atom of the hydroxyl radical, the ACBN0+$U_p({\rm O})$ calculations nearly reproduce the HSE adsorption energies in all four cases (Fig.~\ref{adsorption}b).  Our results suggest the importance of applying self-consistent $U$ corrections to localized states of both adsorbates and substrates.

We then calculate the adsorption energies $E_{\rm ads}$ of HO, NO, and CO on both polar surfaces of In$_2$Se$_3$ by fully optimizing the atomic positions of gas molecules using a unit cell of In$_2$Se$_3$ sheet with its in-plane lattice constants and atomic positions fixed. Specifically, ACBN0+$U_p$ denotes the DFT+$U$ method in which self-consistent Hubbard corrections are applied to both In$_2$Se$_3$ (In-$4d$ and Se-$4p$) and the $2p$ states of gas molecules: $U_p$(O)$=7.74$~eV for HO, $U_p$(O)$=2.47$~eV and $U_p$(N)$=0.95$~eV for NO, $U_p$(O)$=3.61$~eV and $U_p$(C)$=0.20$~eV for CO. Table~\ref{Eads} reports the adsorption energies computed with different methods. We find that for closed-shell molecule CO, the three methods, PBE, ACBN0, and ACBN0+$U_p$ predict similar adsorption strengths that are insensitive to the charge states of polar surfaces. Consistent with our benchmark study, the PBE and ACBN0 values of $E_{\rm ads}$ are much larger (more negative) than ACBN0+$U_p$ for HO adsorbed on both surfaces of In$_2$Se$_3$, with the former two methods indicating a chemical adsorption while the latter method implying a physical adsorption. Interestingly, the adsorption energies for the open-shell molecule NO computed with PBE, ACBN0, and ACBN0+$U_p$ are comparable. We also examine the effects of vdW interactions using the semi-empirical DFT-D3 method. The inclusion of dispersive effects generally increases the binding energies and reduce the adsorption distances. 

\begin{figure}[ht]
\centering
\includegraphics[scale=1.0]{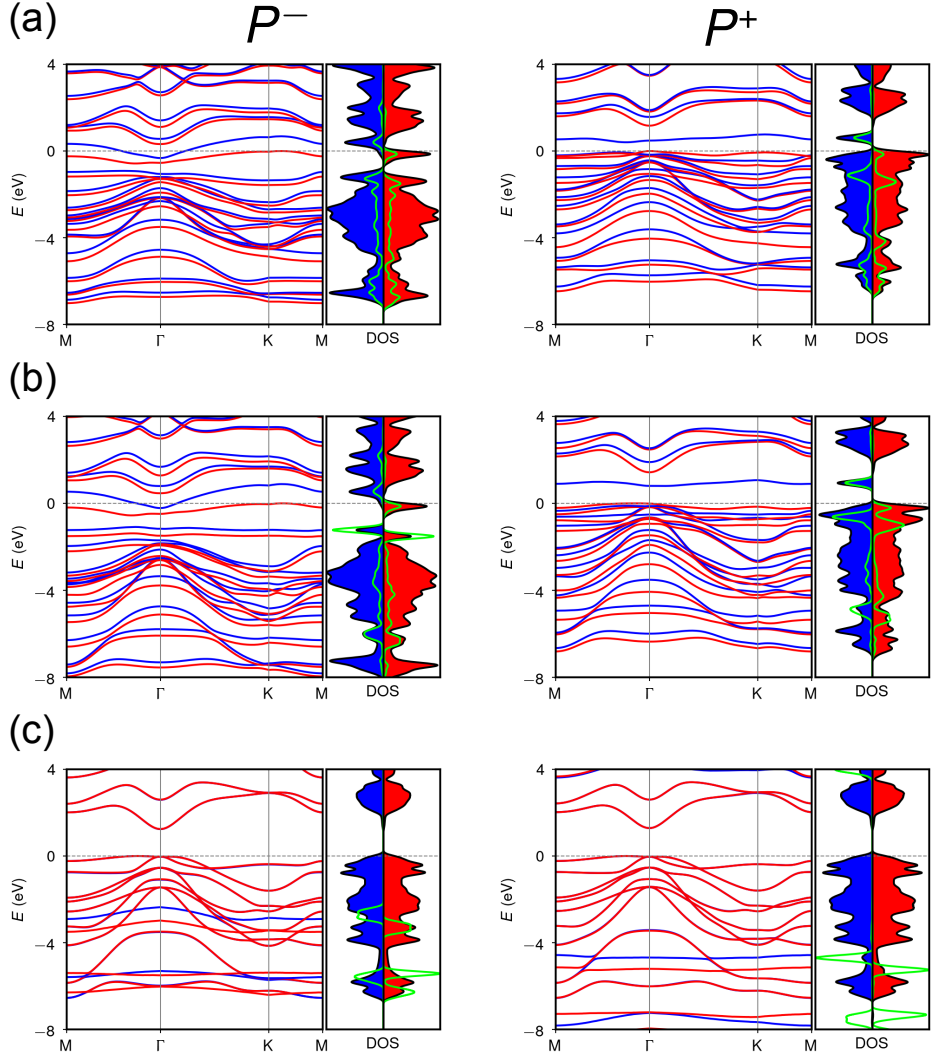}
 \caption{Comparison between the spin-resolved band structures and density of states (DOS) of HO adsorption on $P^-$ (left) and $P^+$ (right) surfaces of $\alpha$-In$_2$Se$_3$ obtained with (a) PBE (b) ACBN0 (c) ACBN0+$U_p$(O). The blue and red solid lines in the band structure represent spin-up and spin-down states, respectively. Filled red and blue curves in the DOS plot represent spin-up and spin down states of In$_2$Se$_3$, the green line denotes O-$2p$ states, respectively.}
  \label{adsorptionHO}
 \end{figure}

The substantially different adsorption strengths of HO predicted by PBE(ACBN0) and ACBN0+$U_p$ can be understood by inspecting the spin-resolved band structures and density of states. As shown in Fig.~\ref{adsorptionHO}, the oxygen $2p$ states hybridize strongly with the states of In$_2$Se$_3$ over a broad energy window within PBE (ACBN0), indicating a strong charge transfer between In$_2$Se$_3$ and HO. Particularly, both PBE and ACBN0 predict a half-filled metallic system for HO adsorbed on the $P^-$ surface. In comparison, the use of $U_p$(O) noticeably downshifts the bands of O-$2p$ characters such that states near $E_f$ remain dominated by In$_2$Se$_3$, consistent with a low $E_{\rm ads}$ and a physisorption character.

\section{Conclusion}
In this work, we investigate the electronic structures of III$_2$-VI$_3$-type 2D FEs with DFT+$U$ and DFT+$U$+$V$ methods using self-consistent Hubbard parameters. Our results show that $U$ values computed with first-principles schemes, DFPT, ACBN0, and eACBN0, are comparable with each other. The self-consistent $U$ is sensitive to the local atomic environment and can potentially serve as a useful local descriptor to differentiate atoms of the same element type in complex materials. The band gaps and band dispersions predicted with ACBN0 are comparable with those calculated with hybrid functional HSE06 for all studied 2D FEs. Importantly, the inclusion of  inter-site Coulomb interaction $V$ is critical for improved descriptions of the electronic structures of vdW heterostructures involving 2D FEs and more covalent 2D materials such as InTe.  We further find that it is important to apply self-consistent $U$ corrections to both adsorbates and substrates to obtain accurate adsorption energies of small molecules on 2D FEs. 

 \section{Acknowledgments}
JH and SL acknowledge the supports from Westlake Foundation and Westlake Multidisciplinary Research Initiative Center. The computational resource is provided by Westlake Supercomputer Center. Y.-W.S. was supported by the NRF of Korea (2017R1A5A1014862, SRC program: vdWMRC Center) and by a KIAS individual grant (CG031509). S.-H.L and Y.-W.S acknowledge the computational support from the CAC of KIAS.

\bibliography{SL}

\begin{thebibliography}{84}%
\makeatletter
\providecommand \@ifxundefined [1]{%
 \@ifx{#1\undefined}
}%
\providecommand \@ifnum [1]{%
 \ifnum #1\expandafter \@firstoftwo
 \else \expandafter \@secondoftwo
 \fi
}%
\providecommand \@ifx [1]{%
 \ifx #1\expandafter \@firstoftwo
 \else \expandafter \@secondoftwo
 \fi
}%
\providecommand \natexlab [1]{#1}%
\providecommand \enquote  [1]{``#1''}%
\providecommand \bibnamefont  [1]{#1}%
\providecommand \bibfnamefont [1]{#1}%
\providecommand \citenamefont [1]{#1}%
\providecommand \href@noop [0]{\@secondoftwo}%
\providecommand \href [0]{\begingroup \@sanitize@url \@href}%
\providecommand \@href[1]{\@@startlink{#1}\@@href}%
\providecommand \@@href[1]{\endgroup#1\@@endlink}%
\providecommand \@sanitize@url [0]{\catcode `\\12\catcode `\$12\catcode
  `\&12\catcode `\#12\catcode `\^12\catcode `\_12\catcode `\%12\relax}%
\providecommand \@@startlink[1]{}%
\providecommand \@@endlink[0]{}%
\providecommand \url  [0]{\begingroup\@sanitize@url \@url }%
\providecommand \@url [1]{\endgroup\@href {#1}{\urlprefix }}%
\providecommand \urlprefix  [0]{URL }%
\providecommand \Eprint [0]{\href }%
\providecommand \doibase [0]{http://dx.doi.org/}%
\providecommand \selectlanguage [0]{\@gobble}%
\providecommand \bibinfo  [0]{\@secondoftwo}%
\providecommand \bibfield  [0]{\@secondoftwo}%
\providecommand \translation [1]{[#1]}%
\providecommand \BibitemOpen [0]{}%
\providecommand \bibitemStop [0]{}%
\providecommand \bibitemNoStop [0]{.\EOS\space}%
\providecommand \EOS [0]{\spacefactor3000\relax}%
\providecommand \BibitemShut  [1]{\csname bibitem#1\endcsname}%
\let\auto@bib@innerbib\@empty
\bibitem [{\citenamefont {Setter}\ \emph {et~al.}(2006)\citenamefont {Setter},
  \citenamefont {Damjanovic}, \citenamefont {Eng}, \citenamefont {Fox},
  \citenamefont {Gevorgian}, \citenamefont {Hong}, \citenamefont {Kingon},
  \citenamefont {Kohlstedt}, \citenamefont {Park}, \citenamefont {Stephenson}
  \emph {et~al.}}]{Setter06p051606}%
  \BibitemOpen
  \bibfield  {author} {\bibinfo {author} {\bibfnamefont {N.}~\bibnamefont
  {Setter}}, \bibinfo {author} {\bibfnamefont {D.}~\bibnamefont {Damjanovic}},
  \bibinfo {author} {\bibfnamefont {L.}~\bibnamefont {Eng}}, \bibinfo {author}
  {\bibfnamefont {G.}~\bibnamefont {Fox}}, \bibinfo {author} {\bibfnamefont
  {S.}~\bibnamefont {Gevorgian}}, \bibinfo {author} {\bibfnamefont
  {S.}~\bibnamefont {Hong}}, \bibinfo {author} {\bibfnamefont {A.}~\bibnamefont
  {Kingon}}, \bibinfo {author} {\bibfnamefont {H.}~\bibnamefont {Kohlstedt}},
  \bibinfo {author} {\bibfnamefont {N.}~\bibnamefont {Park}}, \bibinfo {author}
  {\bibfnamefont {G.}~\bibnamefont {Stephenson}},  \emph {et~al.},\ }\href@noop
  {} {\bibfield  {journal} {\bibinfo  {journal} {J. Appl. Phys.}\ }\textbf
  {\bibinfo {volume} {100}},\ \bibinfo {pages} {051606} (\bibinfo {year}
  {2006})}\BibitemShut {NoStop}%
\bibitem [{\citenamefont {Scott}(2007)}]{Scott07p954}%
  \BibitemOpen
  \bibfield  {author} {\bibinfo {author} {\bibfnamefont {J.~F.}\ \bibnamefont
  {Scott}},\ }\href@noop {} {\bibfield  {journal} {\bibinfo  {journal}
  {Science}\ }\textbf {\bibinfo {volume} {315}},\ \bibinfo {pages} {954}
  (\bibinfo {year} {2007})}\BibitemShut {NoStop}%
\bibitem [{\citenamefont {Huang}\ \emph {et~al.}(2018)\citenamefont {Huang},
  \citenamefont {Zhao}, \citenamefont {Luo}, \citenamefont {Yin}, \citenamefont
  {Lin}, \citenamefont {Hou}, \citenamefont {Tian}, \citenamefont {Duan},\ and\
  \citenamefont {Li}}]{Huang18p1700560}%
  \BibitemOpen
  \bibfield  {author} {\bibinfo {author} {\bibfnamefont {W.}~\bibnamefont
  {Huang}}, \bibinfo {author} {\bibfnamefont {W.}~\bibnamefont {Zhao}},
  \bibinfo {author} {\bibfnamefont {Z.}~\bibnamefont {Luo}}, \bibinfo {author}
  {\bibfnamefont {Y.}~\bibnamefont {Yin}}, \bibinfo {author} {\bibfnamefont
  {Y.}~\bibnamefont {Lin}}, \bibinfo {author} {\bibfnamefont {C.}~\bibnamefont
  {Hou}}, \bibinfo {author} {\bibfnamefont {B.}~\bibnamefont {Tian}}, \bibinfo
  {author} {\bibfnamefont {C.-G.}\ \bibnamefont {Duan}}, \ and\ \bibinfo
  {author} {\bibfnamefont {X.-G.}\ \bibnamefont {Li}},\ }\href {\doibase
  10.1002/aelm.201700560} {\bibfield  {journal} {\bibinfo  {journal} {Adv.
  Electron. Mater.}\ }\textbf {\bibinfo {volume} {4}},\ \bibinfo {pages}
  {1700560} (\bibinfo {year} {2018})}\BibitemShut {NoStop}%
\bibitem [{\citenamefont {Mikolajick}\ \emph {et~al.}(2020)\citenamefont
  {Mikolajick}, \citenamefont {Schroeder},\ and\ \citenamefont
  {Slesazeck}}]{Mikolajick20p1434}%
  \BibitemOpen
  \bibfield  {author} {\bibinfo {author} {\bibfnamefont {T.}~\bibnamefont
  {Mikolajick}}, \bibinfo {author} {\bibfnamefont {U.}~\bibnamefont
  {Schroeder}}, \ and\ \bibinfo {author} {\bibfnamefont {S.}~\bibnamefont
  {Slesazeck}},\ }\href {\doibase 10.1109/ted.2020.2976148} {\bibfield
  {journal} {\bibinfo  {journal} {{IEEE} Trans. Electron Devices}\ }\textbf
  {\bibinfo {volume} {67}},\ \bibinfo {pages} {1434} (\bibinfo {year}
  {2020})}\BibitemShut {NoStop}%
\bibitem [{\citenamefont {Bondurant}(1990)}]{Bondurant90p273}%
  \BibitemOpen
  \bibfield  {author} {\bibinfo {author} {\bibfnamefont {D.}~\bibnamefont
  {Bondurant}},\ }\href {\doibase 10.1080/00150199008008233} {\bibfield
  {journal} {\bibinfo  {journal} {Ferroelectrics}\ }\textbf {\bibinfo {volume}
  {112}},\ \bibinfo {pages} {273} (\bibinfo {year} {1990})}\BibitemShut
  {NoStop}%
\bibitem [{\citenamefont {McAdams}\ \emph {et~al.}(2004)\citenamefont
  {McAdams}, \citenamefont {Acklin}, \citenamefont {Blake}, \citenamefont {Du},
  \citenamefont {Eliason}, \citenamefont {Fong}, \citenamefont {Kraus},
  \citenamefont {Liu}, \citenamefont {Madan}, \citenamefont {Moise},
  \citenamefont {Natarajan}, \citenamefont {Qian}, \citenamefont {Qiu},
  \citenamefont {Remack}, \citenamefont {Rodriguez}, \citenamefont {Roscher},
  \citenamefont {Seshadri},\ and\ \citenamefont {Summerfelt}}]{McAdams04p667}%
  \BibitemOpen
  \bibfield  {author} {\bibinfo {author} {\bibfnamefont {H.}~\bibnamefont
  {McAdams}}, \bibinfo {author} {\bibfnamefont {R.}~\bibnamefont {Acklin}},
  \bibinfo {author} {\bibfnamefont {T.}~\bibnamefont {Blake}}, \bibinfo
  {author} {\bibfnamefont {X.-H.}\ \bibnamefont {Du}}, \bibinfo {author}
  {\bibfnamefont {J.}~\bibnamefont {Eliason}}, \bibinfo {author} {\bibfnamefont
  {J.}~\bibnamefont {Fong}}, \bibinfo {author} {\bibfnamefont {W.}~\bibnamefont
  {Kraus}}, \bibinfo {author} {\bibfnamefont {D.}~\bibnamefont {Liu}}, \bibinfo
  {author} {\bibfnamefont {S.}~\bibnamefont {Madan}}, \bibinfo {author}
  {\bibfnamefont {T.}~\bibnamefont {Moise}}, \bibinfo {author} {\bibfnamefont
  {S.}~\bibnamefont {Natarajan}}, \bibinfo {author} {\bibfnamefont
  {N.}~\bibnamefont {Qian}}, \bibinfo {author} {\bibfnamefont {Y.}~\bibnamefont
  {Qiu}}, \bibinfo {author} {\bibfnamefont {K.}~\bibnamefont {Remack}},
  \bibinfo {author} {\bibfnamefont {J.}~\bibnamefont {Rodriguez}}, \bibinfo
  {author} {\bibfnamefont {J.}~\bibnamefont {Roscher}}, \bibinfo {author}
  {\bibfnamefont {A.}~\bibnamefont {Seshadri}}, \ and\ \bibinfo {author}
  {\bibfnamefont {S.}~\bibnamefont {Summerfelt}},\ }\href {\doibase
  10.1109/jssc.2004.825241} {\bibfield  {journal} {\bibinfo  {journal} {{IEEE}
  J. Solid-State Circuits}\ }\textbf {\bibinfo {volume} {39}},\ \bibinfo
  {pages} {667} (\bibinfo {year} {2004})}\BibitemShut {NoStop}%
\bibitem [{\citenamefont {Onsager}(1944)}]{Onsager44p117}%
  \BibitemOpen
  \bibfield  {author} {\bibinfo {author} {\bibfnamefont {L.}~\bibnamefont
  {Onsager}},\ }\href {\doibase 10.1103/PhysRev.65.117} {\bibfield  {journal}
  {\bibinfo  {journal} {Phys. Rev.}\ }\textbf {\bibinfo {volume} {65}},\
  \bibinfo {pages} {117} (\bibinfo {year} {1944})}\BibitemShut {NoStop}%
\bibitem [{\citenamefont {Wu}\ \emph {et~al.}(2013)\citenamefont {Wu},
  \citenamefont {Burton}, \citenamefont {Tsymbal}, \citenamefont {Zeng},\ and\
  \citenamefont {Jena}}]{Wu13p081406l}%
  \BibitemOpen
  \bibfield  {author} {\bibinfo {author} {\bibfnamefont {M.}~\bibnamefont
  {Wu}}, \bibinfo {author} {\bibfnamefont {J.~D.}\ \bibnamefont {Burton}},
  \bibinfo {author} {\bibfnamefont {E.~Y.}\ \bibnamefont {Tsymbal}}, \bibinfo
  {author} {\bibfnamefont {X.~C.}\ \bibnamefont {Zeng}}, \ and\ \bibinfo
  {author} {\bibfnamefont {P.}~\bibnamefont {Jena}},\ }\href@noop {} {\bibfield
   {journal} {\bibinfo  {journal} {Phys. Rev. B}\ }\textbf {\bibinfo {volume}
  {87}},\ \bibinfo {pages} {081406} (\bibinfo {year} {2013})}\BibitemShut
  {NoStop}%
\bibitem [{\citenamefont {Shirodkar}\ and\ \citenamefont
  {Waghmare}(2014)}]{Shirodkar14p157601}%
  \BibitemOpen
  \bibfield  {author} {\bibinfo {author} {\bibfnamefont {S.~N.}\ \bibnamefont
  {Shirodkar}}\ and\ \bibinfo {author} {\bibfnamefont {U.~V.}\ \bibnamefont
  {Waghmare}},\ }\href@noop {} {\bibfield  {journal} {\bibinfo  {journal}
  {Phys. Rev. Lett.}\ }\textbf {\bibinfo {volume} {112}},\ \bibinfo {pages}
  {157601} (\bibinfo {year} {2014})}\BibitemShut {NoStop}%
\bibitem [{\citenamefont {Bruyer}\ \emph {et~al.}(2016)\citenamefont {Bruyer},
  \citenamefont {Di~Sante}, \citenamefont {Barone}, \citenamefont {Stroppa},
  \citenamefont {Whangbo},\ and\ \citenamefont {Picozzi}}]{Bruyer16p195402}%
  \BibitemOpen
  \bibfield  {author} {\bibinfo {author} {\bibfnamefont {E.}~\bibnamefont
  {Bruyer}}, \bibinfo {author} {\bibfnamefont {D.}~\bibnamefont {Di~Sante}},
  \bibinfo {author} {\bibfnamefont {P.}~\bibnamefont {Barone}}, \bibinfo
  {author} {\bibfnamefont {A.}~\bibnamefont {Stroppa}}, \bibinfo {author}
  {\bibfnamefont {M.-H.}\ \bibnamefont {Whangbo}}, \ and\ \bibinfo {author}
  {\bibfnamefont {S.}~\bibnamefont {Picozzi}},\ }\href@noop {} {\bibfield
  {journal} {\bibinfo  {journal} {Phys. Rev. B}\ }\textbf {\bibinfo {volume}
  {94}},\ \bibinfo {pages} {195402} (\bibinfo {year} {2016})}\BibitemShut
  {NoStop}%
\bibitem [{\citenamefont {Belianinov}\ \emph {et~al.}(2015)\citenamefont
  {Belianinov}, \citenamefont {He}, \citenamefont {Dziaugys}, \citenamefont
  {Maksymovych}, \citenamefont {Eliseev}, \citenamefont {Borisevich},
  \citenamefont {Morozovska}, \citenamefont {Banys}, \citenamefont
  {Vysochanskii},\ and\ \citenamefont {Kalinin}}]{Belianinov15p3808}%
  \BibitemOpen
  \bibfield  {author} {\bibinfo {author} {\bibfnamefont {A.}~\bibnamefont
  {Belianinov}}, \bibinfo {author} {\bibfnamefont {Q.}~\bibnamefont {He}},
  \bibinfo {author} {\bibfnamefont {A.}~\bibnamefont {Dziaugys}}, \bibinfo
  {author} {\bibfnamefont {P.}~\bibnamefont {Maksymovych}}, \bibinfo {author}
  {\bibfnamefont {E.}~\bibnamefont {Eliseev}}, \bibinfo {author} {\bibfnamefont
  {A.}~\bibnamefont {Borisevich}}, \bibinfo {author} {\bibfnamefont
  {A.}~\bibnamefont {Morozovska}}, \bibinfo {author} {\bibfnamefont
  {J.}~\bibnamefont {Banys}}, \bibinfo {author} {\bibfnamefont
  {Y.}~\bibnamefont {Vysochanskii}}, \ and\ \bibinfo {author} {\bibfnamefont
  {S.~V.}\ \bibnamefont {Kalinin}},\ }\href@noop {} {\bibfield  {journal}
  {\bibinfo  {journal} {Nano Lett.}\ }\textbf {\bibinfo {volume} {15}},\
  \bibinfo {pages} {3808} (\bibinfo {year} {2015})}\BibitemShut {NoStop}%
\bibitem [{\citenamefont {Zhou}\ \emph {et~al.}(2017)\citenamefont {Zhou},
  \citenamefont {Wu}, \citenamefont {Zhu}, \citenamefont {Cho}, \citenamefont
  {He}, \citenamefont {Yang}, \citenamefont {Herrera}, \citenamefont {Chu},
  \citenamefont {Han}, \citenamefont {Downer}, \citenamefont {Peng},\ and\
  \citenamefont {Lai}}]{Zhou17p5508}%
  \BibitemOpen
  \bibfield  {author} {\bibinfo {author} {\bibfnamefont {Y.}~\bibnamefont
  {Zhou}}, \bibinfo {author} {\bibfnamefont {D.}~\bibnamefont {Wu}}, \bibinfo
  {author} {\bibfnamefont {Y.}~\bibnamefont {Zhu}}, \bibinfo {author}
  {\bibfnamefont {Y.}~\bibnamefont {Cho}}, \bibinfo {author} {\bibfnamefont
  {Q.}~\bibnamefont {He}}, \bibinfo {author} {\bibfnamefont {X.}~\bibnamefont
  {Yang}}, \bibinfo {author} {\bibfnamefont {K.}~\bibnamefont {Herrera}},
  \bibinfo {author} {\bibfnamefont {Z.}~\bibnamefont {Chu}}, \bibinfo {author}
  {\bibfnamefont {Y.}~\bibnamefont {Han}}, \bibinfo {author} {\bibfnamefont
  {M.~C.}\ \bibnamefont {Downer}}, \bibinfo {author} {\bibfnamefont
  {H.}~\bibnamefont {Peng}}, \ and\ \bibinfo {author} {\bibfnamefont
  {K.}~\bibnamefont {Lai}},\ }\href {\doibase 10.1021/acs.nanolett.7b02198}
  {\bibfield  {journal} {\bibinfo  {journal} {Nano Lett.}\ }\textbf {\bibinfo
  {volume} {17}},\ \bibinfo {pages} {5508} (\bibinfo {year}
  {2017})}\BibitemShut {NoStop}%
\bibitem [{\citenamefont {Chang}\ \emph {et~al.}(2016)\citenamefont {Chang},
  \citenamefont {Liu}, \citenamefont {Lin}, \citenamefont {Wang}, \citenamefont
  {Zhao}, \citenamefont {Zhang}, \citenamefont {Jin}, \citenamefont {Zhong},
  \citenamefont {Hu}, \citenamefont {Duan} \emph {et~al.}}]{Chang16p274}%
  \BibitemOpen
  \bibfield  {author} {\bibinfo {author} {\bibfnamefont {K.}~\bibnamefont
  {Chang}}, \bibinfo {author} {\bibfnamefont {J.}~\bibnamefont {Liu}}, \bibinfo
  {author} {\bibfnamefont {H.}~\bibnamefont {Lin}}, \bibinfo {author}
  {\bibfnamefont {N.}~\bibnamefont {Wang}}, \bibinfo {author} {\bibfnamefont
  {K.}~\bibnamefont {Zhao}}, \bibinfo {author} {\bibfnamefont {A.}~\bibnamefont
  {Zhang}}, \bibinfo {author} {\bibfnamefont {F.}~\bibnamefont {Jin}}, \bibinfo
  {author} {\bibfnamefont {Y.}~\bibnamefont {Zhong}}, \bibinfo {author}
  {\bibfnamefont {X.}~\bibnamefont {Hu}}, \bibinfo {author} {\bibfnamefont
  {W.}~\bibnamefont {Duan}},  \emph {et~al.},\ }\href@noop {} {\bibfield
  {journal} {\bibinfo  {journal} {Science}\ }\textbf {\bibinfo {volume}
  {353}},\ \bibinfo {pages} {274} (\bibinfo {year} {2016})}\BibitemShut
  {NoStop}%
\bibitem [{\citenamefont {Yuan}\ \emph {et~al.}(2019)\citenamefont {Yuan},
  \citenamefont {Luo}, \citenamefont {Chan}, \citenamefont {Xiao},
  \citenamefont {Dai}, \citenamefont {Xie},\ and\ \citenamefont
  {Hao}}]{Yuan19p1775}%
  \BibitemOpen
  \bibfield  {author} {\bibinfo {author} {\bibfnamefont {S.}~\bibnamefont
  {Yuan}}, \bibinfo {author} {\bibfnamefont {X.}~\bibnamefont {Luo}}, \bibinfo
  {author} {\bibfnamefont {H.~L.}\ \bibnamefont {Chan}}, \bibinfo {author}
  {\bibfnamefont {C.}~\bibnamefont {Xiao}}, \bibinfo {author} {\bibfnamefont
  {Y.}~\bibnamefont {Dai}}, \bibinfo {author} {\bibfnamefont {M.}~\bibnamefont
  {Xie}}, \ and\ \bibinfo {author} {\bibfnamefont {J.}~\bibnamefont {Hao}},\
  }\href@noop {} {\bibfield  {journal} {\bibinfo  {journal} {Nat. Commun.}\
  }\textbf {\bibinfo {volume} {10}} (\bibinfo {year} {2019})}\BibitemShut
  {NoStop}%
\bibitem [{\citenamefont {Yang}\ \emph {et~al.}(2018)\citenamefont {Yang},
  \citenamefont {Wu},\ and\ \citenamefont {Li}}]{Yang18p7160}%
  \BibitemOpen
  \bibfield  {author} {\bibinfo {author} {\bibfnamefont {Q.}~\bibnamefont
  {Yang}}, \bibinfo {author} {\bibfnamefont {M.}~\bibnamefont {Wu}}, \ and\
  \bibinfo {author} {\bibfnamefont {J.}~\bibnamefont {Li}},\ }\href@noop {}
  {\bibfield  {journal} {\bibinfo  {journal} {J. Phys. Chem. Lett.}\ }\textbf
  {\bibinfo {volume} {9}},\ \bibinfo {pages} {7160} (\bibinfo {year}
  {2018})}\BibitemShut {NoStop}%
\bibitem [{\citenamefont {Ding}\ \emph {et~al.}(2017)\citenamefont {Ding},
  \citenamefont {Zhu}, \citenamefont {Wang}, \citenamefont {Gao}, \citenamefont
  {Xiao}, \citenamefont {Gu}, \citenamefont {Zhang},\ and\ \citenamefont
  {Zhu}}]{Ding17p14956}%
  \BibitemOpen
  \bibfield  {author} {\bibinfo {author} {\bibfnamefont {W.}~\bibnamefont
  {Ding}}, \bibinfo {author} {\bibfnamefont {J.}~\bibnamefont {Zhu}}, \bibinfo
  {author} {\bibfnamefont {Z.}~\bibnamefont {Wang}}, \bibinfo {author}
  {\bibfnamefont {Y.}~\bibnamefont {Gao}}, \bibinfo {author} {\bibfnamefont
  {D.}~\bibnamefont {Xiao}}, \bibinfo {author} {\bibfnamefont {Y.}~\bibnamefont
  {Gu}}, \bibinfo {author} {\bibfnamefont {Z.}~\bibnamefont {Zhang}}, \ and\
  \bibinfo {author} {\bibfnamefont {W.}~\bibnamefont {Zhu}},\ }\href@noop {}
  {\bibfield  {journal} {\bibinfo  {journal} {Nat. Commun.}\ }\textbf {\bibinfo
  {volume} {8}},\ \bibinfo {pages} {1} (\bibinfo {year} {2017})}\BibitemShut
  {NoStop}%
\bibitem [{\citenamefont {Cui}\ \emph {et~al.}(2018)\citenamefont {Cui},
  \citenamefont {Hu}, \citenamefont {Yan}, \citenamefont {Addiego},
  \citenamefont {Gao}, \citenamefont {Wang}, \citenamefont {Wang},
  \citenamefont {Li}, \citenamefont {Cheng}, \citenamefont {Li}, \citenamefont
  {Zhang}, \citenamefont {Alshareef}, \citenamefont {Wu}, \citenamefont {Zhu},
  \citenamefont {Pan},\ and\ \citenamefont {Li}}]{Cui18p1253}%
  \BibitemOpen
  \bibfield  {author} {\bibinfo {author} {\bibfnamefont {C.}~\bibnamefont
  {Cui}}, \bibinfo {author} {\bibfnamefont {W.-J.}\ \bibnamefont {Hu}},
  \bibinfo {author} {\bibfnamefont {X.}~\bibnamefont {Yan}}, \bibinfo {author}
  {\bibfnamefont {C.}~\bibnamefont {Addiego}}, \bibinfo {author} {\bibfnamefont
  {W.}~\bibnamefont {Gao}}, \bibinfo {author} {\bibfnamefont {Y.}~\bibnamefont
  {Wang}}, \bibinfo {author} {\bibfnamefont {Z.}~\bibnamefont {Wang}}, \bibinfo
  {author} {\bibfnamefont {L.}~\bibnamefont {Li}}, \bibinfo {author}
  {\bibfnamefont {Y.}~\bibnamefont {Cheng}}, \bibinfo {author} {\bibfnamefont
  {P.}~\bibnamefont {Li}}, \bibinfo {author} {\bibfnamefont {X.}~\bibnamefont
  {Zhang}}, \bibinfo {author} {\bibfnamefont {H.~N.}\ \bibnamefont
  {Alshareef}}, \bibinfo {author} {\bibfnamefont {T.}~\bibnamefont {Wu}},
  \bibinfo {author} {\bibfnamefont {W.}~\bibnamefont {Zhu}}, \bibinfo {author}
  {\bibfnamefont {X.}~\bibnamefont {Pan}}, \ and\ \bibinfo {author}
  {\bibfnamefont {L.-J.}\ \bibnamefont {Li}},\ }\href {\doibase
  10.1021/acs.nanolett.7b04852} {\bibfield  {journal} {\bibinfo  {journal}
  {Nano Lett.}\ }\textbf {\bibinfo {volume} {18}},\ \bibinfo {pages} {1253}
  (\bibinfo {year} {2018})}\BibitemShut {NoStop}%
\bibitem [{\citenamefont {Xue}\ \emph {et~al.}(2018{\natexlab{a}})\citenamefont
  {Xue}, \citenamefont {Zhang}, \citenamefont {Hu}, \citenamefont {Hsu},
  \citenamefont {Han}, \citenamefont {Leung}, \citenamefont {Huang},
  \citenamefont {Wan}, \citenamefont {Liu}, \citenamefont {Zhang},
  \citenamefont {He}, \citenamefont {Chang}, \citenamefont {Wang},
  \citenamefont {Zhang},\ and\ \citenamefont {Li}}]{Xue18p4976}%
  \BibitemOpen
  \bibfield  {author} {\bibinfo {author} {\bibfnamefont {F.}~\bibnamefont
  {Xue}}, \bibinfo {author} {\bibfnamefont {J.}~\bibnamefont {Zhang}}, \bibinfo
  {author} {\bibfnamefont {W.}~\bibnamefont {Hu}}, \bibinfo {author}
  {\bibfnamefont {W.-T.}\ \bibnamefont {Hsu}}, \bibinfo {author} {\bibfnamefont
  {A.}~\bibnamefont {Han}}, \bibinfo {author} {\bibfnamefont {S.-F.}\
  \bibnamefont {Leung}}, \bibinfo {author} {\bibfnamefont {J.-K.}\ \bibnamefont
  {Huang}}, \bibinfo {author} {\bibfnamefont {Y.}~\bibnamefont {Wan}}, \bibinfo
  {author} {\bibfnamefont {S.}~\bibnamefont {Liu}}, \bibinfo {author}
  {\bibfnamefont {J.}~\bibnamefont {Zhang}}, \bibinfo {author} {\bibfnamefont
  {J.-H.}\ \bibnamefont {He}}, \bibinfo {author} {\bibfnamefont {W.-H.}\
  \bibnamefont {Chang}}, \bibinfo {author} {\bibfnamefont {Z.~L.}\ \bibnamefont
  {Wang}}, \bibinfo {author} {\bibfnamefont {X.}~\bibnamefont {Zhang}}, \ and\
  \bibinfo {author} {\bibfnamefont {L.-J.}\ \bibnamefont {Li}},\ }\href
  {\doibase 10.1021/acsnano.8b02152} {\bibfield  {journal} {\bibinfo  {journal}
  {{ACS} Nano}\ }\textbf {\bibinfo {volume} {12}},\ \bibinfo {pages} {4976}
  (\bibinfo {year} {2018}{\natexlab{a}})}\BibitemShut {NoStop}%
\bibitem [{\citenamefont {Xiao}\ \emph {et~al.}(2018)\citenamefont {Xiao},
  \citenamefont {Zhu}, \citenamefont {Wang}, \citenamefont {Feng},
  \citenamefont {Hu}, \citenamefont {Dasgupta}, \citenamefont {Han},
  \citenamefont {Wang}, \citenamefont {Muller}, \citenamefont {Martin},
  \citenamefont {Hu},\ and\ \citenamefont {Zhang}}]{Xiao18p227601}%
  \BibitemOpen
  \bibfield  {author} {\bibinfo {author} {\bibfnamefont {J.}~\bibnamefont
  {Xiao}}, \bibinfo {author} {\bibfnamefont {H.}~\bibnamefont {Zhu}}, \bibinfo
  {author} {\bibfnamefont {Y.}~\bibnamefont {Wang}}, \bibinfo {author}
  {\bibfnamefont {W.}~\bibnamefont {Feng}}, \bibinfo {author} {\bibfnamefont
  {Y.}~\bibnamefont {Hu}}, \bibinfo {author} {\bibfnamefont {A.}~\bibnamefont
  {Dasgupta}}, \bibinfo {author} {\bibfnamefont {Y.}~\bibnamefont {Han}},
  \bibinfo {author} {\bibfnamefont {Y.}~\bibnamefont {Wang}}, \bibinfo {author}
  {\bibfnamefont {D.~A.}\ \bibnamefont {Muller}}, \bibinfo {author}
  {\bibfnamefont {L.~W.}\ \bibnamefont {Martin}}, \bibinfo {author}
  {\bibfnamefont {P.}~\bibnamefont {Hu}}, \ and\ \bibinfo {author}
  {\bibfnamefont {X.}~\bibnamefont {Zhang}},\ }\href {\doibase
  10.1103/PhysRevLett.120.227601} {\bibfield  {journal} {\bibinfo  {journal}
  {Phys. Rev. Lett.}\ }\textbf {\bibinfo {volume} {120}},\ \bibinfo {pages}
  {227601} (\bibinfo {year} {2018})}\BibitemShut {NoStop}%
\bibitem [{\citenamefont {Poh}\ \emph {et~al.}(2018)\citenamefont {Poh},
  \citenamefont {Tan}, \citenamefont {Wang}, \citenamefont {Song},
  \citenamefont {Abidi}, \citenamefont {Zhao}, \citenamefont {Dan},
  \citenamefont {Chen}, \citenamefont {Luo}, \citenamefont {Pennycook},
  \citenamefont {Neto},\ and\ \citenamefont {Loh}}]{Poh18p6340}%
  \BibitemOpen
  \bibfield  {author} {\bibinfo {author} {\bibfnamefont {S.~M.}\ \bibnamefont
  {Poh}}, \bibinfo {author} {\bibfnamefont {S.~J.~R.}\ \bibnamefont {Tan}},
  \bibinfo {author} {\bibfnamefont {H.}~\bibnamefont {Wang}}, \bibinfo {author}
  {\bibfnamefont {P.}~\bibnamefont {Song}}, \bibinfo {author} {\bibfnamefont
  {I.~H.}\ \bibnamefont {Abidi}}, \bibinfo {author} {\bibfnamefont
  {X.}~\bibnamefont {Zhao}}, \bibinfo {author} {\bibfnamefont {J.}~\bibnamefont
  {Dan}}, \bibinfo {author} {\bibfnamefont {J.}~\bibnamefont {Chen}}, \bibinfo
  {author} {\bibfnamefont {Z.}~\bibnamefont {Luo}}, \bibinfo {author}
  {\bibfnamefont {S.~J.}\ \bibnamefont {Pennycook}}, \bibinfo {author}
  {\bibfnamefont {A.~H.~C.}\ \bibnamefont {Neto}}, \ and\ \bibinfo {author}
  {\bibfnamefont {K.~P.}\ \bibnamefont {Loh}},\ }\href {\doibase
  10.1021/acs.nanolett.8b02688} {\bibfield  {journal} {\bibinfo  {journal}
  {Nano Lett.}\ }\textbf {\bibinfo {volume} {18}},\ \bibinfo {pages} {6340}
  (\bibinfo {year} {2018})}\BibitemShut {NoStop}%
\bibitem [{\citenamefont {Xue}\ \emph {et~al.}(2018{\natexlab{b}})\citenamefont
  {Xue}, \citenamefont {Hu}, \citenamefont {Lee}, \citenamefont {Lu},
  \citenamefont {Zhang}, \citenamefont {Tang}, \citenamefont {Han},
  \citenamefont {Hsu}, \citenamefont {Tu}, \citenamefont {Chang}, \citenamefont
  {Lien}, \citenamefont {He}, \citenamefont {Zhang}, \citenamefont {Li},\ and\
  \citenamefont {Zhang}}]{Xue18p1803738}%
  \BibitemOpen
  \bibfield  {author} {\bibinfo {author} {\bibfnamefont {F.}~\bibnamefont
  {Xue}}, \bibinfo {author} {\bibfnamefont {W.}~\bibnamefont {Hu}}, \bibinfo
  {author} {\bibfnamefont {K.-C.}\ \bibnamefont {Lee}}, \bibinfo {author}
  {\bibfnamefont {L.-S.}\ \bibnamefont {Lu}}, \bibinfo {author} {\bibfnamefont
  {J.}~\bibnamefont {Zhang}}, \bibinfo {author} {\bibfnamefont {H.-L.}\
  \bibnamefont {Tang}}, \bibinfo {author} {\bibfnamefont {A.}~\bibnamefont
  {Han}}, \bibinfo {author} {\bibfnamefont {W.-T.}\ \bibnamefont {Hsu}},
  \bibinfo {author} {\bibfnamefont {S.}~\bibnamefont {Tu}}, \bibinfo {author}
  {\bibfnamefont {W.-H.}\ \bibnamefont {Chang}}, \bibinfo {author}
  {\bibfnamefont {C.-H.}\ \bibnamefont {Lien}}, \bibinfo {author}
  {\bibfnamefont {J.-H.}\ \bibnamefont {He}}, \bibinfo {author} {\bibfnamefont
  {Z.}~\bibnamefont {Zhang}}, \bibinfo {author} {\bibfnamefont {L.-J.}\
  \bibnamefont {Li}}, \ and\ \bibinfo {author} {\bibfnamefont {X.}~\bibnamefont
  {Zhang}},\ }\href {\doibase 10.1002/adfm.201803738} {\bibfield  {journal}
  {\bibinfo  {journal} {Adv. Funct. Mater.}\ }\textbf {\bibinfo {volume}
  {28}},\ \bibinfo {pages} {1803738} (\bibinfo {year}
  {2018}{\natexlab{b}})}\BibitemShut {NoStop}%
\bibitem [{\citenamefont {Wan}\ \emph {et~al.}(2018)\citenamefont {Wan},
  \citenamefont {Li}, \citenamefont {Li}, \citenamefont {Mao}, \citenamefont
  {Zhu},\ and\ \citenamefont {Zeng}}]{Wan18p14885}%
  \BibitemOpen
  \bibfield  {author} {\bibinfo {author} {\bibfnamefont {S.}~\bibnamefont
  {Wan}}, \bibinfo {author} {\bibfnamefont {Y.}~\bibnamefont {Li}}, \bibinfo
  {author} {\bibfnamefont {W.}~\bibnamefont {Li}}, \bibinfo {author}
  {\bibfnamefont {X.}~\bibnamefont {Mao}}, \bibinfo {author} {\bibfnamefont
  {W.}~\bibnamefont {Zhu}}, \ and\ \bibinfo {author} {\bibfnamefont
  {H.}~\bibnamefont {Zeng}},\ }\href {\doibase 10.1039/c8nr04422h} {\bibfield
  {journal} {\bibinfo  {journal} {Nanoscale}\ }\textbf {\bibinfo {volume}
  {10}},\ \bibinfo {pages} {14885} (\bibinfo {year} {2018})}\BibitemShut
  {NoStop}%
\bibitem [{\citenamefont {Wan}\ \emph {et~al.}(2019)\citenamefont {Wan},
  \citenamefont {Li}, \citenamefont {Li}, \citenamefont {Mao}, \citenamefont
  {Wang}, \citenamefont {Chen}, \citenamefont {Dong}, \citenamefont {Nie},
  \citenamefont {Xiang}, \citenamefont {Liu}, \citenamefont {Zhu},\ and\
  \citenamefont {Zeng}}]{Wan19p1808606}%
  \BibitemOpen
  \bibfield  {author} {\bibinfo {author} {\bibfnamefont {S.}~\bibnamefont
  {Wan}}, \bibinfo {author} {\bibfnamefont {Y.}~\bibnamefont {Li}}, \bibinfo
  {author} {\bibfnamefont {W.}~\bibnamefont {Li}}, \bibinfo {author}
  {\bibfnamefont {X.}~\bibnamefont {Mao}}, \bibinfo {author} {\bibfnamefont
  {C.}~\bibnamefont {Wang}}, \bibinfo {author} {\bibfnamefont {C.}~\bibnamefont
  {Chen}}, \bibinfo {author} {\bibfnamefont {J.}~\bibnamefont {Dong}}, \bibinfo
  {author} {\bibfnamefont {A.}~\bibnamefont {Nie}}, \bibinfo {author}
  {\bibfnamefont {J.}~\bibnamefont {Xiang}}, \bibinfo {author} {\bibfnamefont
  {Z.}~\bibnamefont {Liu}}, \bibinfo {author} {\bibfnamefont {W.}~\bibnamefont
  {Zhu}}, \ and\ \bibinfo {author} {\bibfnamefont {H.}~\bibnamefont {Zeng}},\
  }\href {\doibase 10.1002/adfm.201808606} {\bibfield  {journal} {\bibinfo
  {journal} {Adv. Funct. Mater.}\ }\textbf {\bibinfo {volume} {29}},\ \bibinfo
  {pages} {1808606} (\bibinfo {year} {2019})}\BibitemShut {NoStop}%
\bibitem [{\citenamefont {Perdew}\ and\ \citenamefont
  {Wang}(1992)}]{Perdew92p13244}%
  \BibitemOpen
  \bibfield  {author} {\bibinfo {author} {\bibfnamefont {J.~P.}\ \bibnamefont
  {Perdew}}\ and\ \bibinfo {author} {\bibfnamefont {Y.}~\bibnamefont {Wang}},\
  }\href@noop {} {\bibfield  {journal} {\bibinfo  {journal} {Phys. Rev. B}\
  }\textbf {\bibinfo {volume} {45}},\ \bibinfo {pages} {13244} (\bibinfo {year}
  {1992})}\BibitemShut {NoStop}%
\bibitem [{\citenamefont {Perdew}\ \emph {et~al.}(1996)\citenamefont {Perdew},
  \citenamefont {Burke},\ and\ \citenamefont {Ernzerhof}}]{Perdew96p3865}%
  \BibitemOpen
  \bibfield  {author} {\bibinfo {author} {\bibfnamefont {J.~P.}\ \bibnamefont
  {Perdew}}, \bibinfo {author} {\bibfnamefont {K.}~\bibnamefont {Burke}}, \
  and\ \bibinfo {author} {\bibfnamefont {M.}~\bibnamefont {Ernzerhof}},\ }\href
  {\doibase 10.1103/PhysRevLett.77.3865} {\bibfield  {journal} {\bibinfo
  {journal} {Phys. Rev. Lett.}\ }\textbf {\bibinfo {volume} {77}},\ \bibinfo
  {pages} {3865} (\bibinfo {year} {1996})}\BibitemShut {NoStop}%
\bibitem [{\citenamefont {Mori-S{\'{a}}nchez}\ \emph
  {et~al.}(2006)\citenamefont {Mori-S{\'{a}}nchez}, \citenamefont {Cohen},\
  and\ \citenamefont {Yang}}]{MoriSnchez06p201102}%
  \BibitemOpen
  \bibfield  {author} {\bibinfo {author} {\bibfnamefont {P.}~\bibnamefont
  {Mori-S{\'{a}}nchez}}, \bibinfo {author} {\bibfnamefont {A.~J.}\ \bibnamefont
  {Cohen}}, \ and\ \bibinfo {author} {\bibfnamefont {W.}~\bibnamefont {Yang}},\
  }\href {\doibase 10.1063/1.2403848} {\bibfield  {journal} {\bibinfo
  {journal} {J. Chem. Phys.}\ }\textbf {\bibinfo {volume} {125}},\ \bibinfo
  {pages} {201102} (\bibinfo {year} {2006})}\BibitemShut {NoStop}%
\bibitem [{\citenamefont {Perdew}\ and\ \citenamefont
  {Zunger}(1981)}]{Perdew81p5048}%
  \BibitemOpen
  \bibfield  {author} {\bibinfo {author} {\bibfnamefont {J.~P.}\ \bibnamefont
  {Perdew}}\ and\ \bibinfo {author} {\bibfnamefont {A.}~\bibnamefont
  {Zunger}},\ }\href@noop {} {\bibfield  {journal} {\bibinfo  {journal} {Phys.
  Rev. B}\ }\textbf {\bibinfo {volume} {23}},\ \bibinfo {pages} {5048}
  (\bibinfo {year} {1981})}\BibitemShut {NoStop}%
\bibitem [{\citenamefont {Heyd}\ \emph {et~al.}(2003)\citenamefont {Heyd},
  \citenamefont {Scuseria},\ and\ \citenamefont {Ernzerhof}}]{Heyd03p8207}%
  \BibitemOpen
  \bibfield  {author} {\bibinfo {author} {\bibfnamefont {J.}~\bibnamefont
  {Heyd}}, \bibinfo {author} {\bibfnamefont {G.~E.}\ \bibnamefont {Scuseria}},
  \ and\ \bibinfo {author} {\bibfnamefont {M.}~\bibnamefont {Ernzerhof}},\
  }\href {\doibase 10.1063/1.1564060} {\bibfield  {journal} {\bibinfo
  {journal} {J. Chem. Phys.}\ }\textbf {\bibinfo {volume} {118}},\ \bibinfo
  {pages} {8207} (\bibinfo {year} {2003})}\BibitemShut {NoStop}%
\bibitem [{\citenamefont {Paier}\ \emph {et~al.}(2006)\citenamefont {Paier},
  \citenamefont {Marsman}, \citenamefont {Hummer}, \citenamefont {Kresse},
  \citenamefont {Gerber},\ and\ \citenamefont
  {{\'{A}}ngy{\'{a}}n}}]{Paier06p154709}%
  \BibitemOpen
  \bibfield  {author} {\bibinfo {author} {\bibfnamefont {J.}~\bibnamefont
  {Paier}}, \bibinfo {author} {\bibfnamefont {M.}~\bibnamefont {Marsman}},
  \bibinfo {author} {\bibfnamefont {K.}~\bibnamefont {Hummer}}, \bibinfo
  {author} {\bibfnamefont {G.}~\bibnamefont {Kresse}}, \bibinfo {author}
  {\bibfnamefont {I.~C.}\ \bibnamefont {Gerber}}, \ and\ \bibinfo {author}
  {\bibfnamefont {J.~G.}\ \bibnamefont {{\'{A}}ngy{\'{a}}n}},\ }\href {\doibase
  10.1063/1.2187006} {\bibfield  {journal} {\bibinfo  {journal} {J. Chem.
  Phys.}\ }\textbf {\bibinfo {volume} {124}},\ \bibinfo {pages} {154709}
  (\bibinfo {year} {2006})}\BibitemShut {NoStop}%
\bibitem [{\citenamefont {Marsman}\ \emph {et~al.}(2008)\citenamefont
  {Marsman}, \citenamefont {Paier}, \citenamefont {Stroppa},\ and\
  \citenamefont {Kresse}}]{Marsman08p064201}%
  \BibitemOpen
  \bibfield  {author} {\bibinfo {author} {\bibfnamefont {M.}~\bibnamefont
  {Marsman}}, \bibinfo {author} {\bibfnamefont {J.}~\bibnamefont {Paier}},
  \bibinfo {author} {\bibfnamefont {A.}~\bibnamefont {Stroppa}}, \ and\
  \bibinfo {author} {\bibfnamefont {G.}~\bibnamefont {Kresse}},\ }\href
  {\doibase 10.1088/0953-8984/20/6/064201} {\bibfield  {journal} {\bibinfo
  {journal} {J. Phys.: Condens. Matter}\ }\textbf {\bibinfo {volume} {20}},\
  \bibinfo {pages} {064201} (\bibinfo {year} {2008})}\BibitemShut {NoStop}%
\bibitem [{\citenamefont {Jain}\ \emph {et~al.}(2011)\citenamefont {Jain},
  \citenamefont {Chelikowsky},\ and\ \citenamefont {Louie}}]{Jain11p216806}%
  \BibitemOpen
  \bibfield  {author} {\bibinfo {author} {\bibfnamefont {M.}~\bibnamefont
  {Jain}}, \bibinfo {author} {\bibfnamefont {J.~R.}\ \bibnamefont
  {Chelikowsky}}, \ and\ \bibinfo {author} {\bibfnamefont {S.~G.}\ \bibnamefont
  {Louie}},\ }\href {\doibase 10.1103/PhysRevLett.107.216806} {\bibfield
  {journal} {\bibinfo  {journal} {Phys. Rev. Lett.}\ }\textbf {\bibinfo
  {volume} {107}},\ \bibinfo {pages} {216806} (\bibinfo {year}
  {2011})}\BibitemShut {NoStop}%
\bibitem [{\citenamefont {Fu}\ \emph {et~al.}(2018)\citenamefont {Fu},
  \citenamefont {Sun}, \citenamefont {Luo}, \citenamefont {Li}, \citenamefont
  {Hu},\ and\ \citenamefont {Yang}}]{Fu18p6312}%
  \BibitemOpen
  \bibfield  {author} {\bibinfo {author} {\bibfnamefont {C.-F.}\ \bibnamefont
  {Fu}}, \bibinfo {author} {\bibfnamefont {J.}~\bibnamefont {Sun}}, \bibinfo
  {author} {\bibfnamefont {Q.}~\bibnamefont {Luo}}, \bibinfo {author}
  {\bibfnamefont {X.}~\bibnamefont {Li}}, \bibinfo {author} {\bibfnamefont
  {W.}~\bibnamefont {Hu}}, \ and\ \bibinfo {author} {\bibfnamefont
  {J.}~\bibnamefont {Yang}},\ }\href {\doibase 10.1021/acs.nanolett.8b02561}
  {\bibfield  {journal} {\bibinfo  {journal} {Nano Lett.}\ }\textbf {\bibinfo
  {volume} {18}},\ \bibinfo {pages} {6312} (\bibinfo {year}
  {2018})}\BibitemShut {NoStop}%
\bibitem [{\citenamefont {Anisimov}\ \emph {et~al.}(1991)\citenamefont
  {Anisimov}, \citenamefont {Zaanen},\ and\ \citenamefont
  {Andersen}}]{Anisimov91p943}%
  \BibitemOpen
  \bibfield  {author} {\bibinfo {author} {\bibfnamefont {V.~I.}\ \bibnamefont
  {Anisimov}}, \bibinfo {author} {\bibfnamefont {J.}~\bibnamefont {Zaanen}}, \
  and\ \bibinfo {author} {\bibfnamefont {O.~K.}\ \bibnamefont {Andersen}},\
  }\href@noop {} {\bibfield  {journal} {\bibinfo  {journal} {Phys. Rev. B}\
  }\textbf {\bibinfo {volume} {44}},\ \bibinfo {pages} {943} (\bibinfo {year}
  {1991})}\BibitemShut {NoStop}%
\bibitem [{\citenamefont {Liechtenstein}\ \emph {et~al.}(1995)\citenamefont
  {Liechtenstein}, \citenamefont {Anisimov},\ and\ \citenamefont
  {Zaanen}}]{Liechtenstein95pR5467}%
  \BibitemOpen
  \bibfield  {author} {\bibinfo {author} {\bibfnamefont {A.~I.}\ \bibnamefont
  {Liechtenstein}}, \bibinfo {author} {\bibfnamefont {V.~I.}\ \bibnamefont
  {Anisimov}}, \ and\ \bibinfo {author} {\bibfnamefont {J.}~\bibnamefont
  {Zaanen}},\ }\href {\doibase 10.1103/PhysRevB.52.R5467} {\bibfield  {journal}
  {\bibinfo  {journal} {Phys. Rev. B}\ }\textbf {\bibinfo {volume} {52}},\
  \bibinfo {pages} {R5467} (\bibinfo {year} {1995})}\BibitemShut {NoStop}%
\bibitem [{\citenamefont {Anisimov}\ \emph {et~al.}(1997)\citenamefont
  {Anisimov}, \citenamefont {Aryasetiawan},\ and\ \citenamefont
  {Lichtenstein}}]{Anisimov97p767}%
  \BibitemOpen
  \bibfield  {author} {\bibinfo {author} {\bibfnamefont {V.~I.}\ \bibnamefont
  {Anisimov}}, \bibinfo {author} {\bibfnamefont {F.}~\bibnamefont
  {Aryasetiawan}}, \ and\ \bibinfo {author} {\bibfnamefont {A.}~\bibnamefont
  {Lichtenstein}},\ }\href@noop {} {\bibfield  {journal} {\bibinfo  {journal}
  {J. Phys. Condens. Matter}\ }\textbf {\bibinfo {volume} {9}},\ \bibinfo
  {pages} {767} (\bibinfo {year} {1997})}\BibitemShut {NoStop}%
\bibitem [{\citenamefont {Dudarev}\ \emph {et~al.}(1998)\citenamefont
  {Dudarev}, \citenamefont {Botton}, \citenamefont {Savrasov}, \citenamefont
  {Humphreys},\ and\ \citenamefont {Sutton}}]{Dudarev98p1505}%
  \BibitemOpen
  \bibfield  {author} {\bibinfo {author} {\bibfnamefont {S.}~\bibnamefont
  {Dudarev}}, \bibinfo {author} {\bibfnamefont {G.}~\bibnamefont {Botton}},
  \bibinfo {author} {\bibfnamefont {S.}~\bibnamefont {Savrasov}}, \bibinfo
  {author} {\bibfnamefont {C.}~\bibnamefont {Humphreys}}, \ and\ \bibinfo
  {author} {\bibfnamefont {A.}~\bibnamefont {Sutton}},\ }\href@noop {}
  {\bibfield  {journal} {\bibinfo  {journal} {Phys. Rev. B}\ }\textbf {\bibinfo
  {volume} {57}},\ \bibinfo {pages} {1505} (\bibinfo {year}
  {1998})}\BibitemShut {NoStop}%
\bibitem [{\citenamefont {Kulik}\ \emph {et~al.}(2006)\citenamefont {Kulik},
  \citenamefont {Cococcioni}, \citenamefont {Scherlis},\ and\ \citenamefont
  {Marzari}}]{Kulik06p103001}%
  \BibitemOpen
  \bibfield  {author} {\bibinfo {author} {\bibfnamefont {H.~J.}\ \bibnamefont
  {Kulik}}, \bibinfo {author} {\bibfnamefont {M.}~\bibnamefont {Cococcioni}},
  \bibinfo {author} {\bibfnamefont {D.~A.}\ \bibnamefont {Scherlis}}, \ and\
  \bibinfo {author} {\bibfnamefont {N.}~\bibnamefont {Marzari}},\ }\href@noop
  {} {\bibfield  {journal} {\bibinfo  {journal} {Phys. Rev. Lett.}\ }\textbf
  {\bibinfo {volume} {97}},\ \bibinfo {pages} {103001} (\bibinfo {year}
  {2006})}\BibitemShut {NoStop}%
\bibitem [{\citenamefont {Himmetoglu}\ \emph {et~al.}(2013)\citenamefont
  {Himmetoglu}, \citenamefont {Floris}, \citenamefont {de~Gironcoli},\ and\
  \citenamefont {Cococcioni}}]{Himmetoglu13p14}%
  \BibitemOpen
  \bibfield  {author} {\bibinfo {author} {\bibfnamefont {B.}~\bibnamefont
  {Himmetoglu}}, \bibinfo {author} {\bibfnamefont {A.}~\bibnamefont {Floris}},
  \bibinfo {author} {\bibfnamefont {S.}~\bibnamefont {de~Gironcoli}}, \ and\
  \bibinfo {author} {\bibfnamefont {M.}~\bibnamefont {Cococcioni}},\ }\href
  {\doibase 10.1002/qua.24521} {\bibfield  {journal} {\bibinfo  {journal} {Int.
  J. Quantum Chem.}\ }\textbf {\bibinfo {volume} {114}},\ \bibinfo {pages} {14}
  (\bibinfo {year} {2013})}\BibitemShut {NoStop}%
\bibitem [{\citenamefont {Kulik}\ and\ \citenamefont
  {Marzari}(2008)}]{Kulik08p134314}%
  \BibitemOpen
  \bibfield  {author} {\bibinfo {author} {\bibfnamefont {H.~J.}\ \bibnamefont
  {Kulik}}\ and\ \bibinfo {author} {\bibfnamefont {N.}~\bibnamefont
  {Marzari}},\ }\href {\doibase 10.1063/1.2987444} {\bibfield  {journal}
  {\bibinfo  {journal} {J. Chem. Phys.}\ }\textbf {\bibinfo {volume} {129}},\
  \bibinfo {pages} {134314} (\bibinfo {year} {2008})}\BibitemShut {NoStop}%
\bibitem [{\citenamefont {Cococcioni}\ and\ \citenamefont
  {de~Gironcoli}(2005)}]{Cococcioni05p035105}%
  \BibitemOpen
  \bibfield  {author} {\bibinfo {author} {\bibfnamefont {M.}~\bibnamefont
  {Cococcioni}}\ and\ \bibinfo {author} {\bibfnamefont {S.}~\bibnamefont
  {de~Gironcoli}},\ }\href {\doibase 10.1103/PhysRevB.71.035105} {\bibfield
  {journal} {\bibinfo  {journal} {Phys. Rev. B}\ }\textbf {\bibinfo {volume}
  {71}},\ \bibinfo {pages} {035105} (\bibinfo {year} {2005})}\BibitemShut
  {NoStop}%
\bibitem [{\citenamefont {Springer}\ and\ \citenamefont
  {Aryasetiawan}(1998)}]{Springer98p4364}%
  \BibitemOpen
  \bibfield  {author} {\bibinfo {author} {\bibfnamefont {M.}~\bibnamefont
  {Springer}}\ and\ \bibinfo {author} {\bibfnamefont {F.}~\bibnamefont
  {Aryasetiawan}},\ }\href {\doibase 10.1103/PhysRevB.57.4364} {\bibfield
  {journal} {\bibinfo  {journal} {Phys. Rev. B}\ }\textbf {\bibinfo {volume}
  {57}},\ \bibinfo {pages} {4364} (\bibinfo {year} {1998})}\BibitemShut
  {NoStop}%
\bibitem [{\citenamefont {Kotani}(2000)}]{Kotani00p2413}%
  \BibitemOpen
  \bibfield  {author} {\bibinfo {author} {\bibfnamefont {T.}~\bibnamefont
  {Kotani}},\ }\href {\doibase 10.1088/0953-8984/12/11/307} {\bibfield
  {journal} {\bibinfo  {journal} {J. Phys.: Condens. Matter}\ }\textbf
  {\bibinfo {volume} {12}},\ \bibinfo {pages} {2413} (\bibinfo {year}
  {2000})}\BibitemShut {NoStop}%
\bibitem [{\citenamefont {Aryasetiawan}\ \emph {et~al.}(2004)\citenamefont
  {Aryasetiawan}, \citenamefont {Imada}, \citenamefont {Georges}, \citenamefont
  {Kotliar}, \citenamefont {Biermann},\ and\ \citenamefont
  {Lichtenstein}}]{Aryasetiawan04p195104}%
  \BibitemOpen
  \bibfield  {author} {\bibinfo {author} {\bibfnamefont {F.}~\bibnamefont
  {Aryasetiawan}}, \bibinfo {author} {\bibfnamefont {M.}~\bibnamefont {Imada}},
  \bibinfo {author} {\bibfnamefont {A.}~\bibnamefont {Georges}}, \bibinfo
  {author} {\bibfnamefont {G.}~\bibnamefont {Kotliar}}, \bibinfo {author}
  {\bibfnamefont {S.}~\bibnamefont {Biermann}}, \ and\ \bibinfo {author}
  {\bibfnamefont {A.~I.}\ \bibnamefont {Lichtenstein}},\ }\href {\doibase
  10.1103/PhysRevB.70.195104} {\bibfield  {journal} {\bibinfo  {journal} {Phys.
  Rev. B}\ }\textbf {\bibinfo {volume} {70}},\ \bibinfo {pages} {195104}
  (\bibinfo {year} {2004})}\BibitemShut {NoStop}%
\bibitem [{\citenamefont {Aryasetiawan}\ \emph {et~al.}(2006)\citenamefont
  {Aryasetiawan}, \citenamefont {Karlsson}, \citenamefont {Jepsen},\ and\
  \citenamefont {Sch\"{o}nberger}}]{Aryasetiawan06p125106}%
  \BibitemOpen
  \bibfield  {author} {\bibinfo {author} {\bibfnamefont {F.}~\bibnamefont
  {Aryasetiawan}}, \bibinfo {author} {\bibfnamefont {K.}~\bibnamefont
  {Karlsson}}, \bibinfo {author} {\bibfnamefont {O.}~\bibnamefont {Jepsen}}, \
  and\ \bibinfo {author} {\bibfnamefont {U.}~\bibnamefont {Sch\"{o}nberger}},\
  }\href {\doibase 10.1103/physrevb.74.125106} {\bibfield  {journal} {\bibinfo
  {journal} {Phys. Rev. B}\ }\textbf {\bibinfo {volume} {74}},\ \bibinfo
  {pages} {125106} (\bibinfo {year} {2006})}\BibitemShut {NoStop}%
\bibitem [{\citenamefont {Aichhorn}\ \emph {et~al.}(2009)\citenamefont
  {Aichhorn}, \citenamefont {Pourovskii}, \citenamefont {Vildosola},
  \citenamefont {Ferrero}, \citenamefont {Parcollet}, \citenamefont {Miyake},
  \citenamefont {Georges},\ and\ \citenamefont {Biermann}}]{Aichhorn09p085101}%
  \BibitemOpen
  \bibfield  {author} {\bibinfo {author} {\bibfnamefont {M.}~\bibnamefont
  {Aichhorn}}, \bibinfo {author} {\bibfnamefont {L.}~\bibnamefont
  {Pourovskii}}, \bibinfo {author} {\bibfnamefont {V.}~\bibnamefont
  {Vildosola}}, \bibinfo {author} {\bibfnamefont {M.}~\bibnamefont {Ferrero}},
  \bibinfo {author} {\bibfnamefont {O.}~\bibnamefont {Parcollet}}, \bibinfo
  {author} {\bibfnamefont {T.}~\bibnamefont {Miyake}}, \bibinfo {author}
  {\bibfnamefont {A.}~\bibnamefont {Georges}}, \ and\ \bibinfo {author}
  {\bibfnamefont {S.}~\bibnamefont {Biermann}},\ }\href {\doibase
  10.1103/PhysRevB.80.085101} {\bibfield  {journal} {\bibinfo  {journal} {Phys.
  Rev. B}\ }\textbf {\bibinfo {volume} {80}},\ \bibinfo {pages} {085101}
  (\bibinfo {year} {2009})}\BibitemShut {NoStop}%
\bibitem [{\citenamefont {Miyake}\ \emph {et~al.}(2009)\citenamefont {Miyake},
  \citenamefont {Aryasetiawan},\ and\ \citenamefont
  {Imada}}]{Takashi09p155134}%
  \BibitemOpen
  \bibfield  {author} {\bibinfo {author} {\bibfnamefont {T.}~\bibnamefont
  {Miyake}}, \bibinfo {author} {\bibfnamefont {F.}~\bibnamefont
  {Aryasetiawan}}, \ and\ \bibinfo {author} {\bibfnamefont {M.}~\bibnamefont
  {Imada}},\ }\href {\doibase 10.1103/PhysRevB.80.155134} {\bibfield  {journal}
  {\bibinfo  {journal} {Phys. Rev. B}\ }\textbf {\bibinfo {volume} {80}},\
  \bibinfo {pages} {155134} (\bibinfo {year} {2009})}\BibitemShut {NoStop}%
\bibitem [{\citenamefont {Aichhorn}\ \emph {et~al.}(2010)\citenamefont
  {Aichhorn}, \citenamefont {Biermann}, \citenamefont {Miyake}, \citenamefont
  {Georges},\ and\ \citenamefont {Imada}}]{Aichhorn10p064504}%
  \BibitemOpen
  \bibfield  {author} {\bibinfo {author} {\bibfnamefont {M.}~\bibnamefont
  {Aichhorn}}, \bibinfo {author} {\bibfnamefont {S.}~\bibnamefont {Biermann}},
  \bibinfo {author} {\bibfnamefont {T.}~\bibnamefont {Miyake}}, \bibinfo
  {author} {\bibfnamefont {A.}~\bibnamefont {Georges}}, \ and\ \bibinfo
  {author} {\bibfnamefont {M.}~\bibnamefont {Imada}},\ }\href {\doibase
  10.1103/PhysRevB.82.064504} {\bibfield  {journal} {\bibinfo  {journal} {Phys.
  Rev. B}\ }\textbf {\bibinfo {volume} {82}},\ \bibinfo {pages} {064504}
  (\bibinfo {year} {2010})}\BibitemShut {NoStop}%
\bibitem [{\citenamefont {{\c{S}}a{\c{s}}{\i}o{\u{g}}lu}\ \emph
  {et~al.}(2011)\citenamefont {{\c{S}}a{\c{s}}{\i}o{\u{g}}lu}, \citenamefont
  {Friedrich},\ and\ \citenamefont {Bl\"{u}gel}}]{Sasioglu11p121101}%
  \BibitemOpen
  \bibfield  {author} {\bibinfo {author} {\bibfnamefont {E.}~\bibnamefont
  {{\c{S}}a{\c{s}}{\i}o{\u{g}}lu}}, \bibinfo {author} {\bibfnamefont
  {C.}~\bibnamefont {Friedrich}}, \ and\ \bibinfo {author} {\bibfnamefont
  {S.}~\bibnamefont {Bl\"{u}gel}},\ }\href {\doibase
  10.1103/PhysRevB.83.121101} {\bibfield  {journal} {\bibinfo  {journal} {Phys.
  Rev. B}\ }\textbf {\bibinfo {volume} {83}},\ \bibinfo {pages} {121101}
  (\bibinfo {year} {2011})}\BibitemShut {NoStop}%
\bibitem [{\citenamefont {Timrov}\ \emph {et~al.}(2018)\citenamefont {Timrov},
  \citenamefont {Marzari},\ and\ \citenamefont {Cococcioni}}]{Timrov18p085127}%
  \BibitemOpen
  \bibfield  {author} {\bibinfo {author} {\bibfnamefont {I.}~\bibnamefont
  {Timrov}}, \bibinfo {author} {\bibfnamefont {N.}~\bibnamefont {Marzari}}, \
  and\ \bibinfo {author} {\bibfnamefont {M.}~\bibnamefont {Cococcioni}},\
  }\href@noop {} {\bibfield  {journal} {\bibinfo  {journal} {Phys. Rev. B}\
  }\textbf {\bibinfo {volume} {98}},\ \bibinfo {pages} {085127} (\bibinfo
  {year} {2018})}\BibitemShut {NoStop}%
\bibitem [{\citenamefont {Agapito}\ \emph {et~al.}(2015)\citenamefont
  {Agapito}, \citenamefont {Curtarolo},\ and\ \citenamefont
  {Nardelli}}]{Agapito15p011006}%
  \BibitemOpen
  \bibfield  {author} {\bibinfo {author} {\bibfnamefont {L.~A.}\ \bibnamefont
  {Agapito}}, \bibinfo {author} {\bibfnamefont {S.}~\bibnamefont {Curtarolo}},
  \ and\ \bibinfo {author} {\bibfnamefont {M.~B.}\ \bibnamefont {Nardelli}},\
  }\href@noop {} {\bibfield  {journal} {\bibinfo  {journal} {Phys. Rev. X}\
  }\textbf {\bibinfo {volume} {5}},\ \bibinfo {pages} {011006} (\bibinfo {year}
  {2015})}\BibitemShut {NoStop}%
\bibitem [{\citenamefont {Lee}\ and\ \citenamefont {Son}(2019)}]{Lee19p05967}%
  \BibitemOpen
  \bibfield  {author} {\bibinfo {author} {\bibfnamefont {S.-H.}\ \bibnamefont
  {Lee}}\ and\ \bibinfo {author} {\bibfnamefont {Y.-W.}\ \bibnamefont {Son}},\
  }\href@noop {} {\bibfield  {journal} {\bibinfo  {journal} {arXiv preprint
  arXiv:1911.05967}\ } (\bibinfo {year} {2019})}\BibitemShut {NoStop}%
\bibitem [{\citenamefont {Tancogne-Dejean}\ and\ \citenamefont
  {Rubio}(2019)}]{Tancogne-Dejean19p10813}%
  \BibitemOpen
  \bibfield  {author} {\bibinfo {author} {\bibfnamefont {N.}~\bibnamefont
  {Tancogne-Dejean}}\ and\ \bibinfo {author} {\bibfnamefont {A.}~\bibnamefont
  {Rubio}},\ }\href@noop {} {\bibfield  {journal} {\bibinfo  {journal} {arXiv
  preprint arXiv:1911.10813}\ } (\bibinfo {year} {2019})}\BibitemShut {NoStop}%
\bibitem [{\citenamefont {Anisimov}\ \emph {et~al.}(1996)\citenamefont
  {Anisimov}, \citenamefont {Elfimov}, \citenamefont {Hamada},\ and\
  \citenamefont {Terakura}}]{Anisimov96p4387}%
  \BibitemOpen
  \bibfield  {author} {\bibinfo {author} {\bibfnamefont {V.}~\bibnamefont
  {Anisimov}}, \bibinfo {author} {\bibfnamefont {I.}~\bibnamefont {Elfimov}},
  \bibinfo {author} {\bibfnamefont {N.}~\bibnamefont {Hamada}}, \ and\ \bibinfo
  {author} {\bibfnamefont {K.}~\bibnamefont {Terakura}},\ }\href@noop {}
  {\bibfield  {journal} {\bibinfo  {journal} {Phys. Rev. B}\ }\textbf {\bibinfo
  {volume} {54}},\ \bibinfo {pages} {4387} (\bibinfo {year}
  {1996})}\BibitemShut {NoStop}%
\bibitem [{\citenamefont {Jr}\ and\ \citenamefont
  {Cococcioni}(2010)}]{LeiriaCampoJr10p055602}%
  \BibitemOpen
  \bibfield  {author} {\bibinfo {author} {\bibfnamefont {V.~L.~C.}\
  \bibnamefont {Jr}}\ and\ \bibinfo {author} {\bibfnamefont {M.}~\bibnamefont
  {Cococcioni}},\ }\href {\doibase 10.1088/0953-8984/22/5/055602} {\bibfield
  {journal} {\bibinfo  {journal} {J. Phys.: Condens. Matter}\ }\textbf
  {\bibinfo {volume} {22}},\ \bibinfo {pages} {055602} (\bibinfo {year}
  {2010})}\BibitemShut {NoStop}%
\bibitem [{\citenamefont {Sch{\"u}ler}\ \emph {et~al.}(2013)\citenamefont
  {Sch{\"u}ler}, \citenamefont {R{\"o}sner}, \citenamefont {Wehling},
  \citenamefont {Lichtenstein},\ and\ \citenamefont
  {Katsnelson}}]{Schuler13p036601}%
  \BibitemOpen
  \bibfield  {author} {\bibinfo {author} {\bibfnamefont {M.}~\bibnamefont
  {Sch{\"u}ler}}, \bibinfo {author} {\bibfnamefont {M.}~\bibnamefont
  {R{\"o}sner}}, \bibinfo {author} {\bibfnamefont {T.}~\bibnamefont {Wehling}},
  \bibinfo {author} {\bibfnamefont {A.}~\bibnamefont {Lichtenstein}}, \ and\
  \bibinfo {author} {\bibfnamefont {M.}~\bibnamefont {Katsnelson}},\
  }\href@noop {} {\bibfield  {journal} {\bibinfo  {journal} {Phys. Rev. Lett.}\
  }\textbf {\bibinfo {volume} {111}},\ \bibinfo {pages} {036601} (\bibinfo
  {year} {2013})}\BibitemShut {NoStop}%
\bibitem [{\citenamefont {Wehling}\ \emph {et~al.}(2011)\citenamefont
  {Wehling}, \citenamefont {{\c{S}}a{\c{s}}{\i}o{\u{g}}lu}, \citenamefont
  {Friedrich}, \citenamefont {Lichtenstein}, \citenamefont {Katsnelson},\ and\
  \citenamefont {Bl{\"u}gel}}]{Wehling11p236805}%
  \BibitemOpen
  \bibfield  {author} {\bibinfo {author} {\bibfnamefont {T.}~\bibnamefont
  {Wehling}}, \bibinfo {author} {\bibfnamefont {E.}~\bibnamefont
  {{\c{S}}a{\c{s}}{\i}o{\u{g}}lu}}, \bibinfo {author} {\bibfnamefont
  {C.}~\bibnamefont {Friedrich}}, \bibinfo {author} {\bibfnamefont
  {A.}~\bibnamefont {Lichtenstein}}, \bibinfo {author} {\bibfnamefont
  {M.}~\bibnamefont {Katsnelson}}, \ and\ \bibinfo {author} {\bibfnamefont
  {S.}~\bibnamefont {Bl{\"u}gel}},\ }\href@noop {} {\bibfield  {journal}
  {\bibinfo  {journal} {Phys. Rev. Lett.}\ }\textbf {\bibinfo {volume} {106}},\
  \bibinfo {pages} {236805} (\bibinfo {year} {2011})}\BibitemShut {NoStop}%
\bibitem [{\citenamefont {Hansmann}\ \emph {et~al.}(2013)\citenamefont
  {Hansmann}, \citenamefont {Ayral}, \citenamefont {Vaugier}, \citenamefont
  {Werner},\ and\ \citenamefont {Biermann}}]{Hansmann13p166401}%
  \BibitemOpen
  \bibfield  {author} {\bibinfo {author} {\bibfnamefont {P.}~\bibnamefont
  {Hansmann}}, \bibinfo {author} {\bibfnamefont {T.}~\bibnamefont {Ayral}},
  \bibinfo {author} {\bibfnamefont {L.}~\bibnamefont {Vaugier}}, \bibinfo
  {author} {\bibfnamefont {P.}~\bibnamefont {Werner}}, \ and\ \bibinfo {author}
  {\bibfnamefont {S.}~\bibnamefont {Biermann}},\ }\href@noop {} {\bibfield
  {journal} {\bibinfo  {journal} {Phys. Rev. Lett.}\ }\textbf {\bibinfo
  {volume} {110}},\ \bibinfo {pages} {166401} (\bibinfo {year}
  {2013})}\BibitemShut {NoStop}%
\bibitem [{\citenamefont {Agapito}\ \emph {et~al.}(2013)\citenamefont
  {Agapito}, \citenamefont {Ferretti}, \citenamefont {Calzolari}, \citenamefont
  {Curtarolo},\ and\ \citenamefont {Nardelli}}]{Agapito13p165127}%
  \BibitemOpen
  \bibfield  {author} {\bibinfo {author} {\bibfnamefont {L.~A.}\ \bibnamefont
  {Agapito}}, \bibinfo {author} {\bibfnamefont {A.}~\bibnamefont {Ferretti}},
  \bibinfo {author} {\bibfnamefont {A.}~\bibnamefont {Calzolari}}, \bibinfo
  {author} {\bibfnamefont {S.}~\bibnamefont {Curtarolo}}, \ and\ \bibinfo
  {author} {\bibfnamefont {M.~B.}\ \bibnamefont {Nardelli}},\ }\href@noop {}
  {\bibfield  {journal} {\bibinfo  {journal} {Phys. Rev. B}\ }\textbf {\bibinfo
  {volume} {88}},\ \bibinfo {pages} {165127} (\bibinfo {year}
  {2013})}\BibitemShut {NoStop}%
\bibitem [{\citenamefont {Mosey}\ and\ \citenamefont
  {Carter}(2007)}]{Mosey07p155123}%
  \BibitemOpen
  \bibfield  {author} {\bibinfo {author} {\bibfnamefont {N.~J.}\ \bibnamefont
  {Mosey}}\ and\ \bibinfo {author} {\bibfnamefont {E.~A.}\ \bibnamefont
  {Carter}},\ }\href@noop {} {\bibfield  {journal} {\bibinfo  {journal} {Phys.
  Rev. B}\ }\textbf {\bibinfo {volume} {76}},\ \bibinfo {pages} {155123}
  (\bibinfo {year} {2007})}\BibitemShut {NoStop}%
\bibitem [{\citenamefont {Mosey}\ \emph {et~al.}(2008)\citenamefont {Mosey},
  \citenamefont {Liao},\ and\ \citenamefont {Carter}}]{Mosey08p014103}%
  \BibitemOpen
  \bibfield  {author} {\bibinfo {author} {\bibfnamefont {N.~J.}\ \bibnamefont
  {Mosey}}, \bibinfo {author} {\bibfnamefont {P.}~\bibnamefont {Liao}}, \ and\
  \bibinfo {author} {\bibfnamefont {E.~A.}\ \bibnamefont {Carter}},\
  }\href@noop {} {\bibfield  {journal} {\bibinfo  {journal} {J. Chem. Phys.}\
  }\textbf {\bibinfo {volume} {129}},\ \bibinfo {pages} {014103} (\bibinfo
  {year} {2008})}\BibitemShut {NoStop}%
\bibitem [{\citenamefont {Tancogne-Dejean}\ \emph {et~al.}(2017)\citenamefont
  {Tancogne-Dejean}, \citenamefont {Oliveira},\ and\ \citenamefont
  {Rubio}}]{Tancogne-Dejean17p245133}%
  \BibitemOpen
  \bibfield  {author} {\bibinfo {author} {\bibfnamefont {N.}~\bibnamefont
  {Tancogne-Dejean}}, \bibinfo {author} {\bibfnamefont {M.~J.~T.}\ \bibnamefont
  {Oliveira}}, \ and\ \bibinfo {author} {\bibfnamefont {A.}~\bibnamefont
  {Rubio}},\ }\href {\doibase 10.1103/PhysRevB.96.245133} {\bibfield  {journal}
  {\bibinfo  {journal} {Phys. Rev. B}\ }\textbf {\bibinfo {volume} {96}},\
  \bibinfo {pages} {245133} (\bibinfo {year} {2017})}\BibitemShut {NoStop}%
\bibitem [{\citenamefont {Campo~Jr}\ and\ \citenamefont
  {Cococcioni}(2010)}]{Campo10p055602}%
  \BibitemOpen
  \bibfield  {author} {\bibinfo {author} {\bibfnamefont {V.~L.}\ \bibnamefont
  {Campo~Jr}}\ and\ \bibinfo {author} {\bibfnamefont {M.}~\bibnamefont
  {Cococcioni}},\ }\href@noop {} {\bibfield  {journal} {\bibinfo  {journal} {J.
  Phys.: Condens. Matter}\ }\textbf {\bibinfo {volume} {22}},\ \bibinfo {pages}
  {055602} (\bibinfo {year} {2010})}\BibitemShut {NoStop}%
\bibitem [{\citenamefont {Giannozzi}\ \emph {et~al.}(2009)\citenamefont
  {Giannozzi}, \citenamefont {Baroni}, \citenamefont {Bonini}, \citenamefont
  {Calandra}, \citenamefont {Car}, \citenamefont {Cavazzoni}, \citenamefont
  {Ceresoli}, \citenamefont {Chiarotti}, \citenamefont {Cococcioni},
  \citenamefont {Dabo} \emph {et~al.}}]{Giannozzi09p395502}%
  \BibitemOpen
  \bibfield  {author} {\bibinfo {author} {\bibfnamefont {P.}~\bibnamefont
  {Giannozzi}}, \bibinfo {author} {\bibfnamefont {S.}~\bibnamefont {Baroni}},
  \bibinfo {author} {\bibfnamefont {N.}~\bibnamefont {Bonini}}, \bibinfo
  {author} {\bibfnamefont {M.}~\bibnamefont {Calandra}}, \bibinfo {author}
  {\bibfnamefont {R.}~\bibnamefont {Car}}, \bibinfo {author} {\bibfnamefont
  {C.}~\bibnamefont {Cavazzoni}}, \bibinfo {author} {\bibfnamefont
  {D.}~\bibnamefont {Ceresoli}}, \bibinfo {author} {\bibfnamefont {G.~L.}\
  \bibnamefont {Chiarotti}}, \bibinfo {author} {\bibfnamefont {M.}~\bibnamefont
  {Cococcioni}}, \bibinfo {author} {\bibfnamefont {I.}~\bibnamefont {Dabo}},
  \emph {et~al.},\ }\href@noop {} {\bibfield  {journal} {\bibinfo  {journal}
  {J. Phys. Condens. Matter}\ }\textbf {\bibinfo {volume} {21}},\ \bibinfo
  {pages} {395502} (\bibinfo {year} {2009})}\BibitemShut {NoStop}%
\bibitem [{\citenamefont {Giannozzi}\ \emph {et~al.}(2017)\citenamefont
  {Giannozzi}, \citenamefont {Andreussi}, \citenamefont {Brumme}, \citenamefont
  {Bunau}, \citenamefont {Nardelli}, \citenamefont {Calandra}, \citenamefont
  {Car}, \citenamefont {Cavazzoni}, \citenamefont {Ceresoli}, \citenamefont
  {Cococcioni} \emph {et~al.}}]{Giannozzi17p465901}%
  \BibitemOpen
  \bibfield  {author} {\bibinfo {author} {\bibfnamefont {P.}~\bibnamefont
  {Giannozzi}}, \bibinfo {author} {\bibfnamefont {O.}~\bibnamefont
  {Andreussi}}, \bibinfo {author} {\bibfnamefont {T.}~\bibnamefont {Brumme}},
  \bibinfo {author} {\bibfnamefont {O.}~\bibnamefont {Bunau}}, \bibinfo
  {author} {\bibfnamefont {M.~B.}\ \bibnamefont {Nardelli}}, \bibinfo {author}
  {\bibfnamefont {M.}~\bibnamefont {Calandra}}, \bibinfo {author}
  {\bibfnamefont {R.}~\bibnamefont {Car}}, \bibinfo {author} {\bibfnamefont
  {C.}~\bibnamefont {Cavazzoni}}, \bibinfo {author} {\bibfnamefont
  {D.}~\bibnamefont {Ceresoli}}, \bibinfo {author} {\bibfnamefont
  {M.}~\bibnamefont {Cococcioni}},  \emph {et~al.},\ }\href@noop {} {\bibfield
  {journal} {\bibinfo  {journal} {J. Phys. Condens. Matter}\ }\textbf {\bibinfo
  {volume} {29}},\ \bibinfo {pages} {465901} (\bibinfo {year}
  {2017})}\BibitemShut {NoStop}%
\bibitem [{\citenamefont {Krukau}\ \emph {et~al.}(2006)\citenamefont {Krukau},
  \citenamefont {Vydrov}, \citenamefont {Izmaylov},\ and\ \citenamefont
  {Scuseria}}]{Krukau06p224106}%
  \BibitemOpen
  \bibfield  {author} {\bibinfo {author} {\bibfnamefont {A.~V.}\ \bibnamefont
  {Krukau}}, \bibinfo {author} {\bibfnamefont {O.~A.}\ \bibnamefont {Vydrov}},
  \bibinfo {author} {\bibfnamefont {A.~F.}\ \bibnamefont {Izmaylov}}, \ and\
  \bibinfo {author} {\bibfnamefont {G.~E.}\ \bibnamefont {Scuseria}},\ }\href
  {\doibase 10.1063/1.2404663} {\bibfield  {journal} {\bibinfo  {journal} {J.
  Chem. Phys.}\ }\textbf {\bibinfo {volume} {125}},\ \bibinfo {pages} {224106}
  (\bibinfo {year} {2006})}\BibitemShut {NoStop}%
\bibitem [{\citenamefont {Marzari}\ \emph {et~al.}(2012)\citenamefont
  {Marzari}, \citenamefont {Mostofi}, \citenamefont {Yates}, \citenamefont
  {Souza},\ and\ \citenamefont {Vanderbilt}}]{Marzari12p1419}%
  \BibitemOpen
  \bibfield  {author} {\bibinfo {author} {\bibfnamefont {N.}~\bibnamefont
  {Marzari}}, \bibinfo {author} {\bibfnamefont {A.~A.}\ \bibnamefont
  {Mostofi}}, \bibinfo {author} {\bibfnamefont {J.~R.}\ \bibnamefont {Yates}},
  \bibinfo {author} {\bibfnamefont {I.}~\bibnamefont {Souza}}, \ and\ \bibinfo
  {author} {\bibfnamefont {D.}~\bibnamefont {Vanderbilt}},\ }\href@noop {}
  {\bibfield  {journal} {\bibinfo  {journal} {Rev. Mod. Phys.}\ }\textbf
  {\bibinfo {volume} {84}},\ \bibinfo {pages} {1419} (\bibinfo {year}
  {2012})}\BibitemShut {NoStop}%
\bibitem [{\citenamefont {Pizzi}\ \emph {et~al.}(2020)\citenamefont {Pizzi},
  \citenamefont {Vitale}, \citenamefont {Arita}, \citenamefont {Bl\"{u}gel},
  \citenamefont {Freimuth}, \citenamefont {G{\'{e}}ranton}, \citenamefont
  {Gibertini}, \citenamefont {Gresch}, \citenamefont {Johnson}, \citenamefont
  {Koretsune}, \citenamefont {Iba{\~{n}}ez-Azpiroz}, \citenamefont {Lee},
  \citenamefont {Lihm}, \citenamefont {Marchand}, \citenamefont {Marrazzo},
  \citenamefont {Mokrousov}, \citenamefont {Mustafa}, \citenamefont {Nohara},
  \citenamefont {Nomura}, \citenamefont {Paulatto}, \citenamefont
  {Ponc{\'{e}}}, \citenamefont {Ponweiser}, \citenamefont {Qiao}, \citenamefont
  {Th\"{o}le}, \citenamefont {Tsirkin}, \citenamefont {Wierzbowska},
  \citenamefont {Marzari}, \citenamefont {Vanderbilt}, \citenamefont {Souza},
  \citenamefont {Mostofi},\ and\ \citenamefont {Yates}}]{Pizzi20p165902}%
  \BibitemOpen
  \bibfield  {author} {\bibinfo {author} {\bibfnamefont {G.}~\bibnamefont
  {Pizzi}}, \bibinfo {author} {\bibfnamefont {V.}~\bibnamefont {Vitale}},
  \bibinfo {author} {\bibfnamefont {R.}~\bibnamefont {Arita}}, \bibinfo
  {author} {\bibfnamefont {S.}~\bibnamefont {Bl\"{u}gel}}, \bibinfo {author}
  {\bibfnamefont {F.}~\bibnamefont {Freimuth}}, \bibinfo {author}
  {\bibfnamefont {G.}~\bibnamefont {G{\'{e}}ranton}}, \bibinfo {author}
  {\bibfnamefont {M.}~\bibnamefont {Gibertini}}, \bibinfo {author}
  {\bibfnamefont {D.}~\bibnamefont {Gresch}}, \bibinfo {author} {\bibfnamefont
  {C.}~\bibnamefont {Johnson}}, \bibinfo {author} {\bibfnamefont
  {T.}~\bibnamefont {Koretsune}}, \bibinfo {author} {\bibfnamefont
  {J.}~\bibnamefont {Iba{\~{n}}ez-Azpiroz}}, \bibinfo {author} {\bibfnamefont
  {H.}~\bibnamefont {Lee}}, \bibinfo {author} {\bibfnamefont {J.-M.}\
  \bibnamefont {Lihm}}, \bibinfo {author} {\bibfnamefont {D.}~\bibnamefont
  {Marchand}}, \bibinfo {author} {\bibfnamefont {A.}~\bibnamefont {Marrazzo}},
  \bibinfo {author} {\bibfnamefont {Y.}~\bibnamefont {Mokrousov}}, \bibinfo
  {author} {\bibfnamefont {J.~I.}\ \bibnamefont {Mustafa}}, \bibinfo {author}
  {\bibfnamefont {Y.}~\bibnamefont {Nohara}}, \bibinfo {author} {\bibfnamefont
  {Y.}~\bibnamefont {Nomura}}, \bibinfo {author} {\bibfnamefont
  {L.}~\bibnamefont {Paulatto}}, \bibinfo {author} {\bibfnamefont
  {S.}~\bibnamefont {Ponc{\'{e}}}}, \bibinfo {author} {\bibfnamefont
  {T.}~\bibnamefont {Ponweiser}}, \bibinfo {author} {\bibfnamefont
  {J.}~\bibnamefont {Qiao}}, \bibinfo {author} {\bibfnamefont {F.}~\bibnamefont
  {Th\"{o}le}}, \bibinfo {author} {\bibfnamefont {S.~S.}\ \bibnamefont
  {Tsirkin}}, \bibinfo {author} {\bibfnamefont {M.}~\bibnamefont
  {Wierzbowska}}, \bibinfo {author} {\bibfnamefont {N.}~\bibnamefont
  {Marzari}}, \bibinfo {author} {\bibfnamefont {D.}~\bibnamefont {Vanderbilt}},
  \bibinfo {author} {\bibfnamefont {I.}~\bibnamefont {Souza}}, \bibinfo
  {author} {\bibfnamefont {A.~A.}\ \bibnamefont {Mostofi}}, \ and\ \bibinfo
  {author} {\bibfnamefont {J.~R.}\ \bibnamefont {Yates}},\ }\href {\doibase
  10.1088/1361-648x/ab51ff} {\bibfield  {journal} {\bibinfo  {journal} {J.
  Phys.: Condens. Matter}\ }\textbf {\bibinfo {volume} {32}},\ \bibinfo {pages}
  {165902} (\bibinfo {year} {2020})}\BibitemShut {NoStop}%
\bibitem [{\citenamefont {Supka}\ \emph {et~al.}(2017)\citenamefont {Supka},
  \citenamefont {Lyons}, \citenamefont {Liyanage}, \citenamefont {D'Amico},
  \citenamefont {Orabi}, \citenamefont {Mahatara}, \citenamefont {Gopal},
  \citenamefont {Toher}, \citenamefont {Ceresoli}, \citenamefont {Calzolari},
  \citenamefont {Curtarolo}, \citenamefont {Nardelli},\ and\ \citenamefont
  {Fornari}}]{Supka17p76}%
  \BibitemOpen
  \bibfield  {author} {\bibinfo {author} {\bibfnamefont {A.~R.}\ \bibnamefont
  {Supka}}, \bibinfo {author} {\bibfnamefont {T.~E.}\ \bibnamefont {Lyons}},
  \bibinfo {author} {\bibfnamefont {L.}~\bibnamefont {Liyanage}}, \bibinfo
  {author} {\bibfnamefont {P.}~\bibnamefont {D'Amico}}, \bibinfo {author}
  {\bibfnamefont {R.~A. R.~A.}\ \bibnamefont {Orabi}}, \bibinfo {author}
  {\bibfnamefont {S.}~\bibnamefont {Mahatara}}, \bibinfo {author}
  {\bibfnamefont {P.}~\bibnamefont {Gopal}}, \bibinfo {author} {\bibfnamefont
  {C.}~\bibnamefont {Toher}}, \bibinfo {author} {\bibfnamefont
  {D.}~\bibnamefont {Ceresoli}}, \bibinfo {author} {\bibfnamefont
  {A.}~\bibnamefont {Calzolari}}, \bibinfo {author} {\bibfnamefont
  {S.}~\bibnamefont {Curtarolo}}, \bibinfo {author} {\bibfnamefont {M.~B.}\
  \bibnamefont {Nardelli}}, \ and\ \bibinfo {author} {\bibfnamefont
  {M.}~\bibnamefont {Fornari}},\ }\href {\doibase
  10.1016/j.commatsci.2017.03.055} {\bibfield  {journal} {\bibinfo  {journal}
  {Comp. Mater. Sci.}\ }\textbf {\bibinfo {volume} {136}},\ \bibinfo {pages}
  {76} (\bibinfo {year} {2017})}\BibitemShut {NoStop}%
\bibitem [{\citenamefont {Garrity}\ \emph {et~al.}(2014)\citenamefont
  {Garrity}, \citenamefont {Bennett}, \citenamefont {Rabe},\ and\ \citenamefont
  {Vanderbilt}}]{Garrity14p446}%
  \BibitemOpen
  \bibfield  {author} {\bibinfo {author} {\bibfnamefont {K.~F.}\ \bibnamefont
  {Garrity}}, \bibinfo {author} {\bibfnamefont {J.~W.}\ \bibnamefont
  {Bennett}}, \bibinfo {author} {\bibfnamefont {K.~M.}\ \bibnamefont {Rabe}}, \
  and\ \bibinfo {author} {\bibfnamefont {D.}~\bibnamefont {Vanderbilt}},\
  }\href {\doibase 10.1016/j.commatsci.2013.08.053} {\bibfield  {journal}
  {\bibinfo  {journal} {Comput. Mater. Sci.}\ }\textbf {\bibinfo {volume}
  {81}},\ \bibinfo {pages} {446} (\bibinfo {year} {2014})}\BibitemShut
  {NoStop}%
\bibitem [{\citenamefont {Tiwari}\ \emph {et~al.}(2020)\citenamefont {Tiwari},
  \citenamefont {Birajdar},\ and\ \citenamefont {Ghosh}}]{Tiwari20p235448}%
  \BibitemOpen
  \bibfield  {author} {\bibinfo {author} {\bibfnamefont {R.~P.}\ \bibnamefont
  {Tiwari}}, \bibinfo {author} {\bibfnamefont {B.}~\bibnamefont {Birajdar}}, \
  and\ \bibinfo {author} {\bibfnamefont {R.~K.}\ \bibnamefont {Ghosh}},\ }\href
  {\doibase 10.1103/PhysRevB.101.235448} {\bibfield  {journal} {\bibinfo
  {journal} {Phys. Rev. B}\ }\textbf {\bibinfo {volume} {101}},\ \bibinfo
  {pages} {235448} (\bibinfo {year} {2020})}\BibitemShut {NoStop}%
\bibitem [{\citenamefont {Liu}\ and\ \citenamefont
  {Pantelides}(2019)}]{Liu19p025001}%
  \BibitemOpen
  \bibfield  {author} {\bibinfo {author} {\bibfnamefont {J.}~\bibnamefont
  {Liu}}\ and\ \bibinfo {author} {\bibfnamefont {S.~T.}\ \bibnamefont
  {Pantelides}},\ }\href {\doibase 10.1088/2053-1583/aaf946} {\bibfield
  {journal} {\bibinfo  {journal} {2D Mater.}\ }\textbf {\bibinfo {volume}
  {6}},\ \bibinfo {pages} {025001} (\bibinfo {year} {2019})}\BibitemShut
  {NoStop}%
\bibitem [{\citenamefont {Cohen}(1992)}]{Cohen92p136}%
  \BibitemOpen
  \bibfield  {author} {\bibinfo {author} {\bibfnamefont {R.~E.}\ \bibnamefont
  {Cohen}},\ }\href {\doibase 10.1038/358136a0} {\bibfield  {journal} {\bibinfo
   {journal} {Nature}\ }\textbf {\bibinfo {volume} {358}},\ \bibinfo {pages}
  {136} (\bibinfo {year} {1992})}\BibitemShut {NoStop}%
\bibitem [{\citenamefont {Lee}\ and\ \citenamefont {Kim}(2012)}]{Lee12p781}%
  \BibitemOpen
  \bibfield  {author} {\bibinfo {author} {\bibfnamefont {W.-J.}\ \bibnamefont
  {Lee}}\ and\ \bibinfo {author} {\bibfnamefont {Y.-S.}\ \bibnamefont {Kim}},\
  }\href {\doibase 10.3938/jkps.60.781} {\bibfield  {journal} {\bibinfo
  {journal} {J. Korean Phys. Soc.}\ }\textbf {\bibinfo {volume} {60}},\
  \bibinfo {pages} {781} (\bibinfo {year} {2012})}\BibitemShut {NoStop}%
\bibitem [{\citenamefont {Cococcioni}\ and\ \citenamefont
  {Marzari}(2019)}]{Cococcioni19p033801}%
  \BibitemOpen
  \bibfield  {author} {\bibinfo {author} {\bibfnamefont {M.}~\bibnamefont
  {Cococcioni}}\ and\ \bibinfo {author} {\bibfnamefont {N.}~\bibnamefont
  {Marzari}},\ }\href {\doibase 10.1103/physrevmaterials.3.033801} {\bibfield
  {journal} {\bibinfo  {journal} {Phys. Rev. Mater.}\ }\textbf {\bibinfo
  {volume} {3}},\ \bibinfo {pages} {033801} (\bibinfo {year}
  {2019})}\BibitemShut {NoStop}%
\bibitem [{\citenamefont {Duong}\ \emph {et~al.}(2017)\citenamefont {Duong},
  \citenamefont {Yun},\ and\ \citenamefont {Lee}}]{Duong17p11803}%
  \BibitemOpen
  \bibfield  {author} {\bibinfo {author} {\bibfnamefont {D.~L.}\ \bibnamefont
  {Duong}}, \bibinfo {author} {\bibfnamefont {S.~J.}\ \bibnamefont {Yun}}, \
  and\ \bibinfo {author} {\bibfnamefont {Y.~H.}\ \bibnamefont {Lee}},\ }\href
  {\doibase 10.1021/acsnano.7b07436} {\bibfield  {journal} {\bibinfo  {journal}
  {{ACS} Nano}\ }\textbf {\bibinfo {volume} {11}},\ \bibinfo {pages} {11803}
  (\bibinfo {year} {2017})}\BibitemShut {NoStop}%
\bibitem [{\citenamefont {Chhowalla}\ \emph {et~al.}(2016)\citenamefont
  {Chhowalla}, \citenamefont {Jena},\ and\ \citenamefont
  {Zhang}}]{Chhowalla16p16052}%
  \BibitemOpen
  \bibfield  {author} {\bibinfo {author} {\bibfnamefont {M.}~\bibnamefont
  {Chhowalla}}, \bibinfo {author} {\bibfnamefont {D.}~\bibnamefont {Jena}}, \
  and\ \bibinfo {author} {\bibfnamefont {H.}~\bibnamefont {Zhang}},\ }\href
  {\doibase 10.1038/natrevmats.2016.52} {\bibfield  {journal} {\bibinfo
  {journal} {Nat. Rev. Mater.}\ }\textbf {\bibinfo {volume} {1}},\ \bibinfo
  {pages} {16052} (\bibinfo {year} {2016})}\BibitemShut {NoStop}%
\bibitem [{\citenamefont {Li}\ \emph {et~al.}(2014)\citenamefont {Li},
  \citenamefont {Li},\ and\ \citenamefont {Yang}}]{Li14p018301}%
  \BibitemOpen
  \bibfield  {author} {\bibinfo {author} {\bibfnamefont {X.}~\bibnamefont
  {Li}}, \bibinfo {author} {\bibfnamefont {Z.}~\bibnamefont {Li}}, \ and\
  \bibinfo {author} {\bibfnamefont {J.}~\bibnamefont {Yang}},\ }\href {\doibase
  10.1103/physrevlett.112.018301} {\bibfield  {journal} {\bibinfo  {journal}
  {Phys. Rev. Lett.}\ }\textbf {\bibinfo {volume} {112}},\ \bibinfo {pages}
  {018301} (\bibinfo {year} {2014})}\BibitemShut {NoStop}%
\bibitem [{\citenamefont {Grimme}(2006)}]{Grimme06p1787}%
  \BibitemOpen
  \bibfield  {author} {\bibinfo {author} {\bibfnamefont {S.}~\bibnamefont
  {Grimme}},\ }\href {\doibase 10.1002/jcc.20495} {\bibfield  {journal}
  {\bibinfo  {journal} {J. Comput. Chem.}\ }\textbf {\bibinfo {volume} {27}},\
  \bibinfo {pages} {1787} (\bibinfo {year} {2006})}\BibitemShut {NoStop}%
\bibitem [{\citenamefont {Levchenko}\ and\ \citenamefont
  {Rappe}(2008)}]{Levchenko08p256101}%
  \BibitemOpen
  \bibfield  {author} {\bibinfo {author} {\bibfnamefont {S.~V.}\ \bibnamefont
  {Levchenko}}\ and\ \bibinfo {author} {\bibfnamefont {A.~M.}\ \bibnamefont
  {Rappe}},\ }\href@noop {} {\bibfield  {journal} {\bibinfo  {journal} {Phys.
  Rev. Lett.}\ }\textbf {\bibinfo {volume} {100}} (\bibinfo {year}
  {2008})}\BibitemShut {NoStop}%
\bibitem [{\citenamefont {Kolpak}\ \emph {et~al.}(2008)\citenamefont {Kolpak},
  \citenamefont {Li}, \citenamefont {Shao}, \citenamefont {Rappe},\ and\
  \citenamefont {Bonnell}}]{Kolpak08p036102}%
  \BibitemOpen
  \bibfield  {author} {\bibinfo {author} {\bibfnamefont {A.~M.}\ \bibnamefont
  {Kolpak}}, \bibinfo {author} {\bibfnamefont {D.}~\bibnamefont {Li}}, \bibinfo
  {author} {\bibfnamefont {R.}~\bibnamefont {Shao}}, \bibinfo {author}
  {\bibfnamefont {A.~M.}\ \bibnamefont {Rappe}}, \ and\ \bibinfo {author}
  {\bibfnamefont {D.~A.}\ \bibnamefont {Bonnell}},\ }\href@noop {} {\bibfield
  {journal} {\bibinfo  {journal} {Phys. Rev. Lett.}\ }\textbf {\bibinfo
  {volume} {101}} (\bibinfo {year} {2008})}\BibitemShut {NoStop}%
\bibitem [{\citenamefont {Kakekhani}\ \emph {et~al.}(2016)\citenamefont
  {Kakekhani}, \citenamefont {Ismail-Beigi},\ and\ \citenamefont
  {Altman}}]{Kakekhani16p302}%
  \BibitemOpen
  \bibfield  {author} {\bibinfo {author} {\bibfnamefont {A.}~\bibnamefont
  {Kakekhani}}, \bibinfo {author} {\bibfnamefont {S.}~\bibnamefont
  {Ismail-Beigi}}, \ and\ \bibinfo {author} {\bibfnamefont {E.~I.}\
  \bibnamefont {Altman}},\ }\href {\doibase 10.1016/j.susc.2015.10.055}
  {\bibfield  {journal} {\bibinfo  {journal} {Surf. Sci.}\ }\textbf {\bibinfo
  {volume} {650}},\ \bibinfo {pages} {302} (\bibinfo {year}
  {2016})}\BibitemShut {NoStop}%
\bibitem [{\citenamefont {Kakekhani}\ and\ \citenamefont
  {Ismail-Beigi}(2016)}]{Kakekhani16p19676}%
  \BibitemOpen
  \bibfield  {author} {\bibinfo {author} {\bibfnamefont {A.}~\bibnamefont
  {Kakekhani}}\ and\ \bibinfo {author} {\bibfnamefont {S.}~\bibnamefont
  {Ismail-Beigi}},\ }\href {\doibase 10.1039/c6cp03170f} {\bibfield  {journal}
  {\bibinfo  {journal} {Phys. Chem. Chem. Phys.}\ }\textbf {\bibinfo {volume}
  {18}},\ \bibinfo {pages} {19676} (\bibinfo {year} {2016})}\BibitemShut
  {NoStop}%
\bibitem [{\citenamefont {Tang}\ \emph {et~al.}(2020)\citenamefont {Tang},
  \citenamefont {Shang}, \citenamefont {Gu}, \citenamefont {Du},\ and\
  \citenamefont {Kou}}]{Tang20p7331}%
  \BibitemOpen
  \bibfield  {author} {\bibinfo {author} {\bibfnamefont {X.}~\bibnamefont
  {Tang}}, \bibinfo {author} {\bibfnamefont {J.}~\bibnamefont {Shang}},
  \bibinfo {author} {\bibfnamefont {Y.}~\bibnamefont {Gu}}, \bibinfo {author}
  {\bibfnamefont {A.}~\bibnamefont {Du}}, \ and\ \bibinfo {author}
  {\bibfnamefont {L.}~\bibnamefont {Kou}},\ }\href {\doibase
  10.1039/d0ta00854k} {\bibfield  {journal} {\bibinfo  {journal} {J. Mater.
  Chem.A}\ }\textbf {\bibinfo {volume} {8}},\ \bibinfo {pages} {7331} (\bibinfo
  {year} {2020})}\BibitemShut {NoStop}%
\bibitem [{\citenamefont {Mason}\ \emph {et~al.}(2004)\citenamefont {Mason},
  \citenamefont {Grinberg},\ and\ \citenamefont {Rappe}}]{Mason04p161401}%
  \BibitemOpen
  \bibfield  {author} {\bibinfo {author} {\bibfnamefont {S.~E.}\ \bibnamefont
  {Mason}}, \bibinfo {author} {\bibfnamefont {I.}~\bibnamefont {Grinberg}}, \
  and\ \bibinfo {author} {\bibfnamefont {A.~M.}\ \bibnamefont {Rappe}},\ }\href
  {\doibase 10.1103/PhysRevB.69.161401} {\bibfield  {journal} {\bibinfo
  {journal} {Phys. Rev. B}\ }\textbf {\bibinfo {volume} {69}},\ \bibinfo
  {pages} {161401} (\bibinfo {year} {2004})}\BibitemShut {NoStop}%
\end{thebibliography}%

\newpage
\begin{table*}[ht]
\centering
\caption{Comparison of on-site Hubbard $U$ values (in eV) and the corresponding band gaps for $\alpha$-In$_2$Se$_3$. The out-of-plane polarization points from Se1 to Se3 as shown in Figure 1. The LR-cDFT values obtained with a unit cell are given in square brackets.}
\begin{ruledtabular}
\begin{tabular}{c|ccc|ccc}
& \multicolumn{3}{c|}{\{ $U_d$(In), $U_p$(Se) \} } & \multicolumn{3}{c}{ \{ $U_p$(In), $U_p$(Se) \}  } \\
\hline
Atom & LR-cDFT & DFPT & ACBN0  & LR-cDFT & DFPT & ACBN0 \\
\hline
Se1 & 5.32 [7.83]         & 3.85   & 3.83       & 5.32 [7.83]          & 3.91   & 3.79    \\
In1  & -                 &12.81  & 15.40     & 0.78 [0.86]          & 1.26    &0.02    \\
Se2 & 5.66 [5.86]       &3.44   & 3.56       & 5.66 [5.86]          &3.42     &3.52    \\
In2  & -                 &13.89 &15.32      &0.78 [0.80]           & 1.05    & 0.02    \\
Se3 & 5.86 [8.40]      &3.71    & 3.11       & 5.86 [8.40]           &3.72     & 3.07   \\
\hline
$E_g$ &  1.57 [1.94]       & 1.34 & 1.32        & 1.57 [1.94]         & 1.33      & 1.30    \\ 
\hline
 \multicolumn{7}{c}{$E_g${(PBE)} = 0.80 }\\
\hline
 \multicolumn{7}{c}{$E_g${(HSE06)} = 1.48 }
\end{tabular}
\end{ruledtabular}
\label{Uvalues}
\end{table*}

\newpage
\begin{table}[ht]
\centering
\caption{Self-consistent on-site Hubbard $U$ values for Se and In $p$-orbitals 
and inter-site Hubbard $V_{\alpha\beta}$ values (in eV) between $\alpha$- and $\beta$-orbitals ($\alpha,\beta=s,p$)
belong to the nearest neighboring atoms of $\alpha$-In$_2$Se$_3$ 
obtained from eACBN0 method. The out-of-plane polarization points from Se1 to Se3.
The notation for $U$ follows Table~\ref{Uvalues}.
The table for $V_{\alpha\beta}$ can be read as follows:
the inter-site $V_{ss}$ between the $s$-orbital of In1 atom (denoted as In1-$s$) and $s$-orbital of Se1
atoms (denoted as Se1-$s$) is 0.84 eV and so on.
}
\begin{ruledtabular}
\begin{tabular}{c|ccccc}
\multirow{2}{*} {\{ $U_p$(In), $U_p$(Se) \} }  & Se1  & In1 &  Se2   &   In2 &  Se3  \\
                                 &  4.06  &  0.11  &  3.58  & 0.12  &3.14 \\
\hline
\hline
\multirow{6}{*} {$V_{ss}$, $V_{sp}$, $V_{pp}$} &                & Se1-$s$  & Se1-$p$  & Se2-$s$ & Se2-$p$    \\
                           & In1-$s$   &  0.84        &   1.66       &   0.78      &   1.52        \\   
                           & In1-$p$   & 0.64        &    1.70       &   0.61      &   1.59          \\
\cline{2-6}                           
                           &              & Se2-$s$ & Se2-$p$  & Se3-$s$  &  Se3-$p$   \\
                           &In2-$s$   &   0.90      &    1.66       &   0.84     &     1.53   \\
                          & In2-$p$    &   0.64      &     1.68      &   0.57     &     1.61  
\end{tabular}
\end{ruledtabular}
\label{UVvalues}
\end{table}

\newpage
 \begin{table*}[ht]
\centering
\caption{Self-consistent on-site Hubbard $U$ values (in eV) from ACBN0 for III$_2$-VI$_3$ compounds (III=Al, Ga, In; VI=S, Se, Te). Hubbard corrections are applied to the $p$-states of Al, S, Se, and Te, and to the $d$-states of Ga and In, respectively. The out-of-plane polarization points from VI1 to VI3.}
\begin{ruledtabular}
\begin{tabular}{c|cccccccc}
Atom & Al$_2$S$_3$ & Al$_2$Se$_3$ & Al$_2$Te$_3$ & Ga$_2$S$_3$ & Ga$_2$Se$_3$ & Ga$_2$Te$_3$ & In$_2$S$_3$ & In$_2$Te$_3$ \\
\hline
VI1 & 4.73 & 4.01 & 3.23 &  4.42 &  3.76 &  2.51 &  4.62 &    3.18 \\
III1 & 0.01 & 0.02 & 0.05 & 19.93 & 20.06 & 20.20 & 15.28 &  15.60 \\
VI2 & 4.36 & 3.72 & 3.03 &  4.19 &  3.60 &  2.90 &  4.26 &    3.01 \\
III2 & 0.01 & 0.02 & 0.06 & 19.81 & 19.98 & 20.20 & 15.16 &  15.57 \\
VI3  & 3.85 & 3.24 & 2.38 &  3.48 &  2.99 &  2.53 &  3.71 &    2.45\\

\end{tabular}
\end{ruledtabular}
\label{Uvaluesall}
\end{table*}

\newpage
\begin{table*}[ht]
\centering
\caption{Polarization-dependent adsorption energies in eV for small molecules on $\alpha$-In$_2$Se$_3$ for 1.0 monolayer coverage. The adsorption distances in \AA~are given in square brackets. }
\begin{ruledtabular}
\begin{tabular}{ccccccc}
\multirow{2}{*}{Method}&\multicolumn{2}{c}{HO} & \multicolumn{2}{c}{NO} & \multicolumn{2}{c}{CO} \\
\cline{2-3}\cline{4-5}\cline{6-7}
& $P^+$ & $P^-$ & $P^+$ & $P^-$ & $P^+$ & $P^-$ \\
\hline
PBE & $-0.794$~[1.65] & $-0.968$~[1.62] & $-0.019$~[3.53]& $-0.047$~[3.05]& $-0.008$~[3.72]& $-0.011$~[3.74]  \\
ACBN0 & $-0.558$~[1.74] & $-0.720$~[1.63] &$-0.022$~[3.53] & $-0.083$~[3.19] &$-0.009$~[3.72] & $-0.012$~[3.54]  \\
ACBN0+$U_p$ & $-0.019$~[3.32]& $-0.011$~[3.67]& $-0.069$~[3.41]&$-0.119$~[3.21] &$-0.009$~[3.74] & $-0.009$~[3.47] \\
PBE+$\rm D3$ & $-0.885$~[1.65] & $-1.066$~[1.57] & $-0.079$~[3.27]& $-0.119$~[2.93]& $-0.071$~[3.66]& $-0.087$~[3.29]  \\
ACBN0+$\rm D3$ & $-0.650$~[1.69] & $-0.816$~[1.58] &$-0.083$~[3.31] & $-0.154$~[3.07] &$-0.073$~[3.66] & $-0.093$~[3.20]  \\
ACBN0+$U_p$+$\rm D3$ & $-0.080$~[2.97]& $-0.051$~[3.62]& $-0.131$~[3.27]&$-0.188$~[3.07] &$-0.089$~[3.32] & $-0.113$~[2.86] \\
\end{tabular}
\end{ruledtabular}
\label{Eads}
\end{table*}

\newpage
 
\newpage

\newpage
 \clearpage

\newpage
 
 \newpage
 
 \newpage

  \newpage
 
   \newpage

\end{document}